\pdfoutput=1

\documentclass[11pt,twoside,a4paper,cmspaper,final,collab]{cms-tdr}
\def\svnVersion{249991}\def\cmsCernNo{2014-015}\def\cmsCernDate{\today}\def\cmsMessage{Published in the Journal of High Energy Physics as \href{http://dx.doi.org/10.1007/JHEP06(2014)055}{\doi{10.1007/JHEP06(2014)055.}}}

\begin{document}\cmsNoteHeader{SUS-13-012}

\hyphenation{had-ron-i-za-tion}
\hyphenation{cal-or-i-me-ter}
\hyphenation{de-vices}

\RCS$Revision: 234776 $
\RCS$HeadURL: svn+ssh://svn.cern.ch/reps/tdr2/papers/SUS-13-012/trunk/SUS-13-012.tex $
\RCS$Id: SUS-13-012.tex 234776 2014-04-02 01:49:36Z hatake $
\newcommand{\fulllumi}{19.5\xspace}
\providecommand{\ZZ}{\cPZ\cPZ\xspace}
\providecommand{\WW}{\PW\PW\xspace}
\providecommand{\WZ}{\PW\cPZ\xspace}
\providecommand\qqbar{\cPq\cPaq\xspace}

\providecommand{\MHT}{\ensuremath{H_{\mathrm{T}}\kern -1.2em/\kern0.48 em}\xspace}
\newcommand{\njets}{\ensuremath{N_{\text{Jets}}}\xspace}
\newcommand\wpj{{W+jets}\xspace}
\newcommand{\znunu}{\ensuremath{\cPZ \to \cPgn \cPagn}\xspace}
\newcommand{\znunubr}{\ensuremath{\cPZ(\cPgn \cPagn )}\xspace}
\newcommand{\zmumubr}{\ensuremath{\cPZ(\Pgmp\Pgmm)}\xspace}
\newcommand{\zellellbr}{\ensuremath{\cPZ ( \ell^{+} \ell^{-} )}\xspace}
\newcommand{\wlnubr}{\ensuremath{\PW (\ell \cPgn )}\xspace}
\newcommand{\wtaunubr}{\ensuremath{\PW (\Pgt \cPgn )}\xspace}
\providecommand{\tauh}{\ensuremath{\tau_\mathrm{h}}\xspace}
\newcommand{\RZgamma}{\ensuremath{R_{\cPZ/\gamma} }\xspace}
\newcommand{\RZmumugamma}{\ensuremath{R_{\cPZ(\Pgmp\Pgmm)/\gamma} }\xspace}
\providecommand{\mt}{\ensuremath{m_\mathrm{T}}\xspace}
\newcommand{\lsp}{\PSGczDo\xspace}

\cmsNoteHeader{SUS-13-012} 
\title{Search for new physics in the multijet and missing transverse momentum final state in proton-proton collisions at $\sqrt{s}=8$\TeV}

\author{The CMS Collaboration}

\date{\today}

\abstract{A search for new physics is performed in multijet events with large missing
transverse momentum produced in proton-proton collisions at $\sqrt{s}=8$\TeV
using a data sample corresponding to an integrated luminosity of 19.5\fbinv collected with
the CMS detector at the LHC. The data sample is divided into
three jet multiplicity categories (3--5, 6--7, and ${\ge}8$ jets), and studied
further in bins of two variables: the scalar sum of jet transverse
momenta and the missing transverse momentum. The observed numbers of events
in various categories are consistent with backgrounds expected from standard model processes.
Exclusion limits are presented for several simplified
supersymmetric models of squark or gluino pair production.}

\hypersetup{%
pdfauthor={CMS Collaboration},%
pdftitle={Search for new physics in the multijet and missing transverse momentum final state in proton-proton collisions at sqrt(s) = 8 TeV},%
pdfsubject={CMS},%
pdfkeywords={CMS, physics, SUSY, hadronic, BSM}}

\maketitle 

\section{Introduction}
\label{sec:introduction}

The standard model of particle physics (SM) successfully describes a wide
variety of observations in high energy physics. The recent discovery of a new
scalar boson with a mass of about 125\GeV~\cite{ATLAS-Higgs,CMS-Higgs,CMS-Higgs-Long}
at the CERN Large Hadron Collider (LHC) marks another success for the
SM, as its properties measured so far are consistent with those
of the long-sought Higgs boson. However, its mass is predicted to be unstable
against quadratically divergent quantum-loop corrections, which
suggests the presence of physics beyond the SM.
Supersymmetry (SUSY) is a well-explored extension that
addresses various shortcomings of the SM.
SUSY postulates a new symmetry, relating
fermionic and bosonic degrees of freedom, and introduces a superpartner for
each SM particle. Radiative corrections due to SUSY particles can compensate
the contribution of the SM particles and thereby stabilize the mass of the Higgs boson.
In $R$-parity-conserving models~\cite{FarrarEtAl}, SUSY particles are produced in pairs,
and the lightest SUSY particle (LSP) is stable. If weakly interacting and neutral,
the LSP is a potential dark matter candidate.

This paper reports an inclusive search for physics beyond the SM in multijet
events with large missing transverse momentum produced in pp collisions at a
centre-of-mass energy $\sqrt{s}=8$\TeV at the LHC.
The data sample used corresponds to an integrated luminosity of
$\fulllumi\fbinv$ collected by the Compact Muon Solenoid (CMS) experiment~\cite{CMS}.
This final state is motivated by many extensions of the
SM, for example those given in Refs.~\cite{SUSY0,LittleHiggs,UED}.
At the LHC, both the CMS and ATLAS collaborations have performed
SUSY searches in all-hadronic final states~\cite{RA2,RA2_2011,MT2_2011,RA1_2012,RA2b_2012,ATLASJetMET2011,ATLASMultiJet2011,ATLASMultiJet2012,ATLASBJetMET2012}.
For all these searches, the observed numbers of events were consistent with
the expected SM background, and exclusion limits were set
in the context of the constrained minimal supersymmetric extension of the standard
model (CMSSM)~\cite{Chamseddine:1982jx,Arnowitt:1992aq,CMSSM} and
various simplified models~\cite{Alwall:2008ag,SMSPaper}.
Contrary to the CMSSM case, the masses of particles are free
parameters in simplified models,
thus allowing a generic study of the parameter space of SUSY and SUSY-like theories.
Simplified models of squark and gluino pair production are used
to interpret the search results in this paper.

This analysis follows previous
inclusive searches~\cite{RA2,RA2_2011} that require at least three jets in
the final state.
These searches are most sensitive to the hypothetical
production of pairs of squarks and gluinos, where the
squarks (gluinos) each decay to one (two) jets and an undetected LSP.
We extend the analyses of Refs.~\cite{RA2,RA2_2011} by subdividing
the data into three exclusive jet multiplicity categories: \njets = 3--5, 6--7, and $\geq$8,
which renders the analysis more sensitive to a variety
of final-state topologies resulting from longer cascades
of squarks and gluinos, and hence in a larger number of jets.
The search regions with higher jet multiplicities extend the sensitivity
of the analysis to models in which the gluino often decays into top quarks.
While other analyses exploit the presence of bottom-quark jets in signal
events to discriminate against background \cite{RA1_2012, RA2b_2012}, this
analysis follows a complementary strategy by requiring a large number of jets,
thus helping to keep the signal efficiency for fully hadronic final states as high as possible.

The events in each jet multiplicity category are further divided according to
variables that characterize the total visible hadronic activity (\HT) and
the momentum imbalance
(\MHT) in an event, both defined in
the plane transverse to the beam.
Due to the presence of a number of energetic jets and two LSPs in the final
state, the signal events are expected to have large \HT and \MHT.
The main SM processes contributing to this final state are
\cPZ+jets events, where the \cPZ\ boson decays to a pair of neutrinos
($\znunubr$+jets), and \PW+jets and $\ttbar$ events, where a \PW\ boson decays
to an $\Pe$, $\Pgm$, or $\tau$ lepton ($\wlnubr$+jets). The presence of at least one
neutrino in these events provides a source of genuine \MHT. Another background
category is quantum chromodynamics (QCD) multijet events with
large \MHT from leptonic decays of heavy-flavour hadrons
inside the jets, jet energy mismeasurement, or instrumental noise and
non-functioning detector components. All these backgrounds are determined
using the data, with as little reliance on simulation as possible.

\section{The CMS detector and event reconstruction}
\label{sec:detector}
The CMS detector is a multipurpose apparatus, described in detail in Ref.~\cite{CMS}.
The CMS coordinate system is defined with the origin at the centre of the detector and
the $z$ axis along the anticlockwise beam direction. The polar angle $\theta$ is measured
with respect to the $z$ axis, and the azimuthal angle $\phi$ (measured in radians) in the plane perpendicular
to that axis.
Charged-particle trajectories are measured with a silicon pixel and strip tracker, covering
$\abs{\eta} < 2.5$, where the pseudorapidity $\eta$ is defined as $\eta = -\ln [\tan (\theta/2)]$.
Immersed in the $3.8\unit{T}$ magnetic field provided by a 6\unit{m} diameter superconducting solenoid,
which also encircles the calorimeters,
the tracking system provides transverse momentum (\pt) resolution of approximately
1.5\% for charged particles with $\pt\sim100$\GeV.
A lead-tungstate crystal electromagnetic calorimeter and a brass-and-scintillator hadron
calorimeter surround the tracking volume and cover the region $\abs{\eta} < 3$. Steel and
quartz-fibre hadron forward calorimeters extend the coverage to $\abs{\eta}\le 5$.
Muons are identified in gas ionization detectors embedded in the steel flux return yoke of the magnet.
The events used for this search are recorded using a two-level trigger system described in Ref.~\cite{CMS}.

The recorded events are required to have at least one well-identified
interaction vertex with $z$ position within 24\unit{cm} from the nominal centre
of the detector and transverse distance from the $z$ axis less than 2\unit{cm}.
The primary vertex is the one with the largest sum of \pt-squared
of all the associated tracks, and is assumed to correspond to the hard-scattering process.
The events are reconstructed using a particle-flow (PF) algorithm~\cite{PFT-09-001}.
This algorithm reconstructs a list of particles in each event,
namely charged and neutral hadrons, photons, muons, and electrons,
combining the information from the tracker, the calorimeters, and the muon system.
These particles are then clustered into jets using the anti-\kt clustering algorithm~\cite{antikt}
with a size parameter of 0.5.
Contributions from additional pp collisions overlapping with the event of interest
(pileup) are mitigated
by discarding charged particles not associated with the primary vertex and using
the Fastjet tools~\cite{PU_JET_AREAS,JET_AREAS} to account for the neutral pileup component.
Corrections to jet energy are applied to
account for the variation of the response in \pt and $\eta$ ~\cite{JETJINST}.
Missing transverse momentum (\ETslash{}) is reconstructed as magnitude of the vector
sum of \pt of all the reconstructed PF particles~\cite{METJINST,METPAS}.

\section{Sample selection}
\label{sec:eventselection}

The search regions are first defined using a loose baseline
selection with the following requirements:

\begin{itemize}
\item $\njets \geq 3$, where \njets is
      the number of jets with $\pt > 50\GeV$ and $| \eta | < 2.5$.
\item $\HT> 500\GeV$, with  $\HT = \sum_{\text{jets}} \pt$, where the sum
      includes all jets with $\pt > 50\GeV$ and $\abs{\eta} < 2.5$.
\item $\MHT> 200\GeV$, with $\MHT = \abs{\vec{\MHT}} = \abs{-\sum_{\text{jets}} \ptvec}$,
      where in this case, jets are required to satisfy $\pt>30 \GeV$ and $\abs{\eta}<5$.
\item $\abs{\Delta\phi(\ptvec^{\text{Jet1}}, \vec{\MHT})} > 0.5$,
      $\abs{\Delta\phi(\ptvec^{\text{Jet2}}, \vec{\MHT})} > 0.5$,  and
      $\abs{\Delta\phi(\ptvec^{\text{Jet3}}, \vec{\MHT})} > 0.3$,
      vetoing the events
      where $\vec{\MHT}$ is aligned with one of the three highest \pt jets.
      This requirement rejects most of the QCD multijet events in which a single
      mismeasured jet yields high \MHT.
\item Events containing isolated muons or electrons with $\pt>10$ \GeV are
      vetoed in order to reject \ttbar and $\PW/\cPZ$+jets events with leptons
      in the final state.
      Both the $\Pe$ and $\Pgm$ are required to produce a good quality track that
      is matched to the primary interaction vertex~\cite{EGMPAS,MUORECO}. The isolation is measured
      as the scalar \pt sum of PF particles ($\pt^\text{sum}$),
      except the lepton itself,
      within a cone of width $\Delta R=\sqrt{\smash[b]{(\Delta\eta)^2+(\Delta\phi)^2}} =  0.3$
      for $\Pe$ (0.4 for $\Pgm$) around the lepton.
      The $\pt^\text{sum}$ is required to be less than 20\% (15\%) of the \pt of the $\Pe$ ($\Pgm$).
\item In addition, events affected by instrumental effects, particles from
      non-collision sources, or poorly reconstructed kinematic variables are rejected (event
      cleaning)~\cite{METJINST,METPAS}.
      Events are also rejected if a jet with  $\pt>30\GeV$ has
      more than 95\% of its energy from PF photon candidates
      or more than 90\% from PF neutral hadron candidates.
\end{itemize}

The data sample used for this analysis was collected using trigger algorithms
that required events to have $\HT > 350$\GeV and $\ETslash> 100$\GeV. The trigger
efficiencies are measured to be greater than 99\% for the offline baseline selection of
$\HT > 500$\GeV and $\MHT > 200$\GeV in all jet multiplicity categories
used in this search.
A sample of 11\,753 events is selected after applying the baseline criteria. The selected
events are divided into 36 non-overlapping search regions defined in terms of \njets, \HT, and \MHT,
as listed in the first three columns of Table~\ref{tab:FinalEventYields}.

Several Monte Carlo (MC) simulation samples are used to model the signal as well as to
develop and validate the background estimation methods. The
\ttbar, \PW/\cPZ+jets, $\gamma$+jets, and QCD multijet background samples are produced using the
\MADGRAPH{}5~\cite{madgraph} generator at leading order (LO), interfaced with the
\PYTHIA 6.4.24~\cite{pythia} parton-shower model, and scaled to the
next-to-leading order (NLO) or next-to-next-to-leading
order cross section predictions~\cite{Kidonakis:2010dk,Melnikov:2006kv}.
The events are processed
through a \GEANTfour simulation of the detector~\cite{Geant4-2}. The SUSY signal samples are generated using \MADGRAPH{}5,
the CTEQ6L~\cite{CTEQ6} parton distribution functions (PDF), and
are simulated using the CMS fast simulation package~\cite{Fastsim}. The underlying event description used
for the MC simulated samples is described in Ref.~\cite{pythia-z2}. The effect of pileup interactions
is included by adding a number of simulated
minimum bias events, on top of the hard interaction, to match the distribution observed in data.

\section{Background estimation}
\label{sec:bg}

In this search, all backgrounds are measured from data
using methods similar to those described in Refs.~\cite{RA2,RA2_2011}.
The $\znunubr$+jets background is estimated using $\gamma$+jets events, exploiting
their electroweak correspondence to \cPZ+jets production for boson \pt above $\sim$100\GeV.
The \cPZ+jets and $\gamma$+jets events exhibit
similar characteristics, apart from electroweak coupling differences and asymptotically
vanishing residual mass effects.
The \ttbar or \wlnubr{}+jets events satisfy the search selection when the $\Pe$/$\Pgm$
is not identified or isolated, or is out of the detector acceptance (``lost-lepton'' background)
or when a $\tau$ lepton decays hadronically
($\tauh$ background). The lost-lepton background is estimated by reweighting events in
a $\Pgm$+jets data control sample with measured
lepton efficiencies. The estimation of the $\tauh$ background starts
from a similar $\Pgm$+jets sample, replacing the muon with a jet
sampled as a function of jet \pt from $\tauh$ templates obtained from simulation.
The QCD multijet background is measured using a ``rebalance-and-smear" method~\cite{RA2,RA2_2011}.
The kinematical characteristics of multijet events are predicted from
data by applying a fitting procedure that imposes zero missing
transverse momentum on each event, and then smearing the jets
according to data-corrected jet energy resolution values.
The relative contribution of the various backgrounds
varies in the different search regions.
\subsection{Estimation of \texorpdfstring{$\znunubr$}{Z(nu nu)} +jets background}

Photons and \cPZ\ bosons exhibit similar kinematic properties at high \pt,
and therefore the hadronic component of an event containing either a high-\pt photon
or Z boson is similar~\cite{Bern:2011pa,Kuhn:2005gv,Ask:2011xf,Bern:2012pa}.
The $\gamma$+jets sample used to evaluate the \znunubr{}+jets event rate is collected by triggering on
events with a $\gamma$ candidate and large \HT.
The photon candidates are reconstructed using the energy deposited in
the electromagnetic calorimeter~\cite{GAMMAPAS,EGMRECO}.
Photon candidates with $\pt>100\GeV$ and $\abs{\eta}< 1.44$ or $1.566 <\abs{\eta}<2.5$
are used in this analysis, and are required to have their lateral shower
profile consistent with that of a photon produced in the hard-scattering process (a prompt photon).
To veto electrons misidentified as photons, the candidates with an
associated track in the pixel detector are rejected.
A photon candidate is required to satisfy tight isolation requirements
based on the sum over \pt values of the PF candidates that lie within
a cone of radius $\Delta R = 0.3$ around the direction of its momentum.

The contribution to the $\gamma$+jets control sample from events in which the photon candidate
originates from the misidentification of jet fragments (background photons) is measured using a template
method, which exploits the difference between the shower profile of prompt (signal)
and background photons, using the distribution of a modified
second moment of the electromagnetic energy cluster around
its mean $\eta$ position~\cite{GAMMAPAS}. The distribution
(template) for background events is obtained
from a sideband region defined by selecting photons that satisfy
very loose photon identification and isolation requirements but
fail the stringent isolation requirements. The distribution for
signal events is obtained from simulation. The sum of the two
templates is fit to the observed distribution, with the normalization
(background and signal yields) of each template determined in the fit.
On average, 93\% of selected $\gamma$+jets candidate events
are determined to originate from prompt photons.

\begin{figure}[!htbp]
  \centering
   \includegraphics[width=0.45\textwidth]{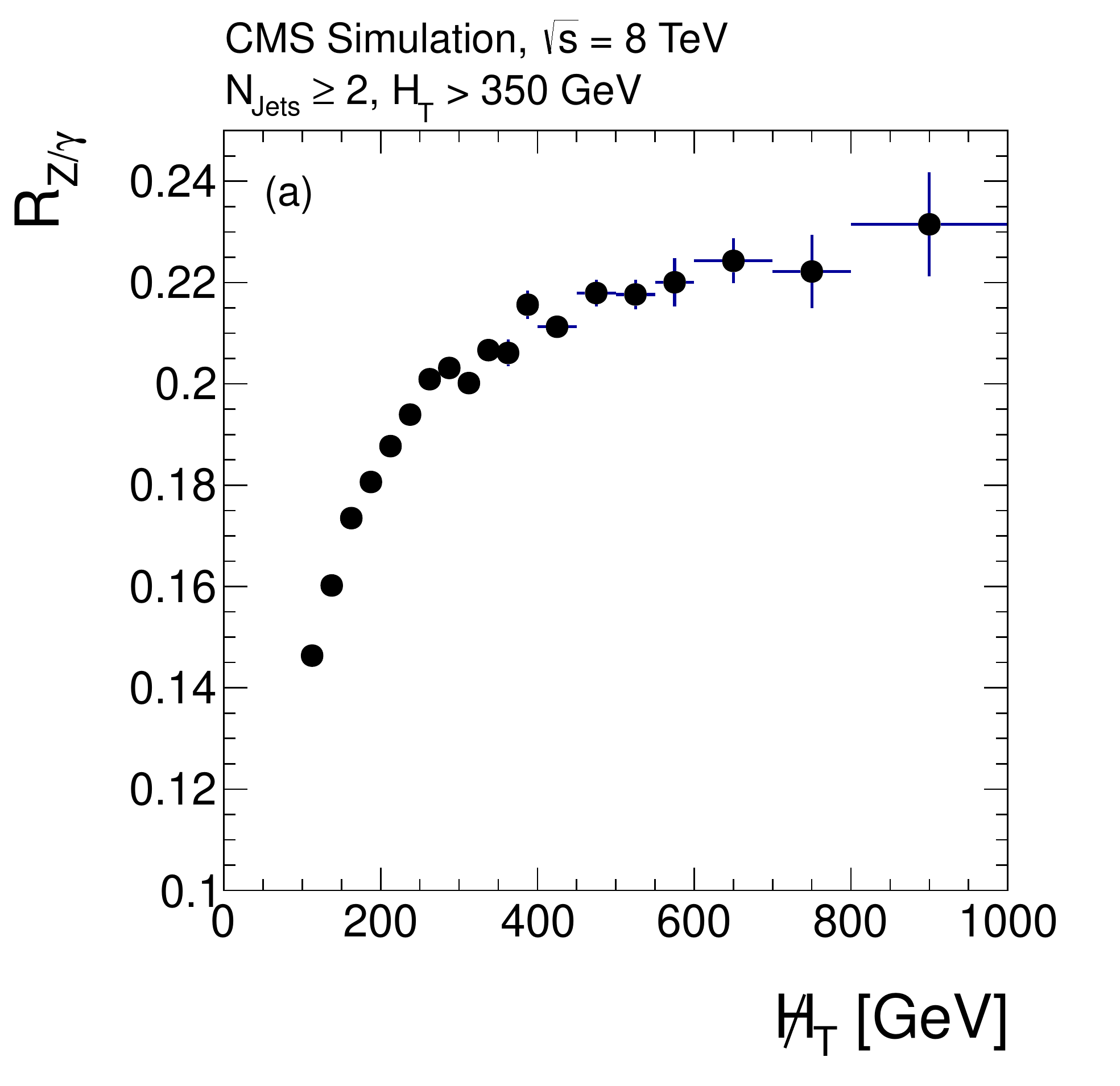}
   \includegraphics[width=0.45\textwidth]{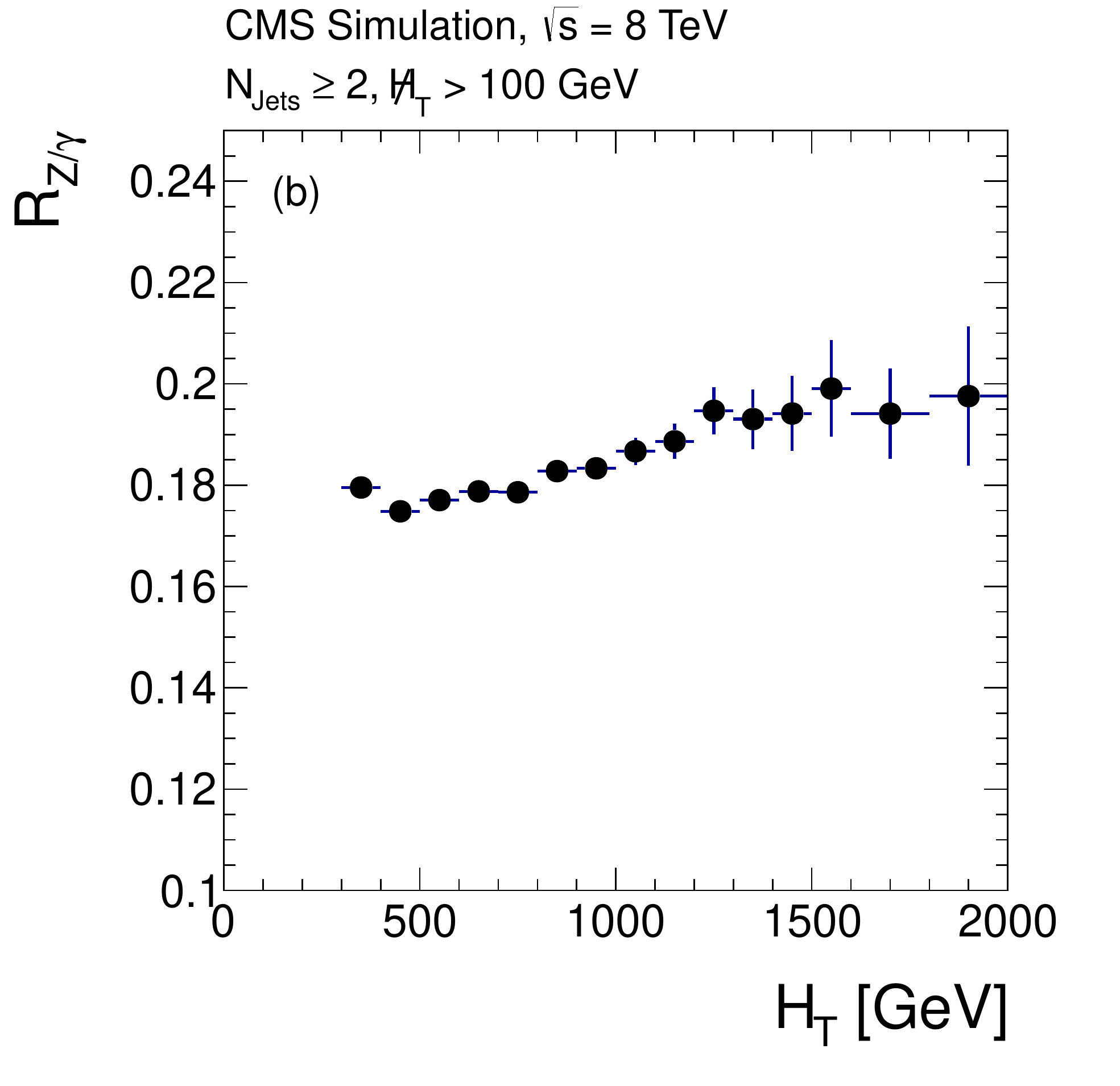} \\
   \includegraphics[width=0.45\textwidth]{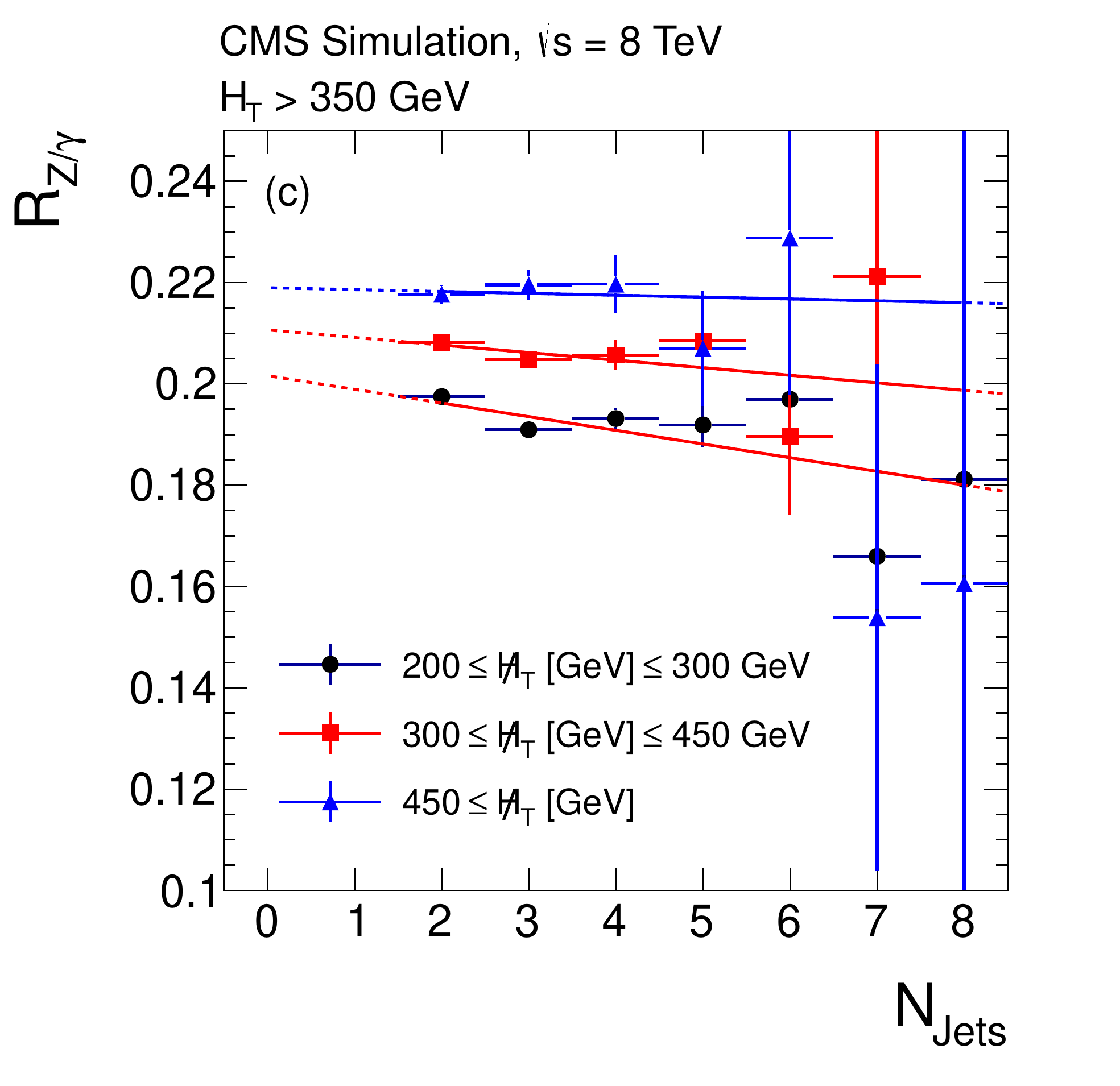}
   \includegraphics[width=0.45\textwidth]{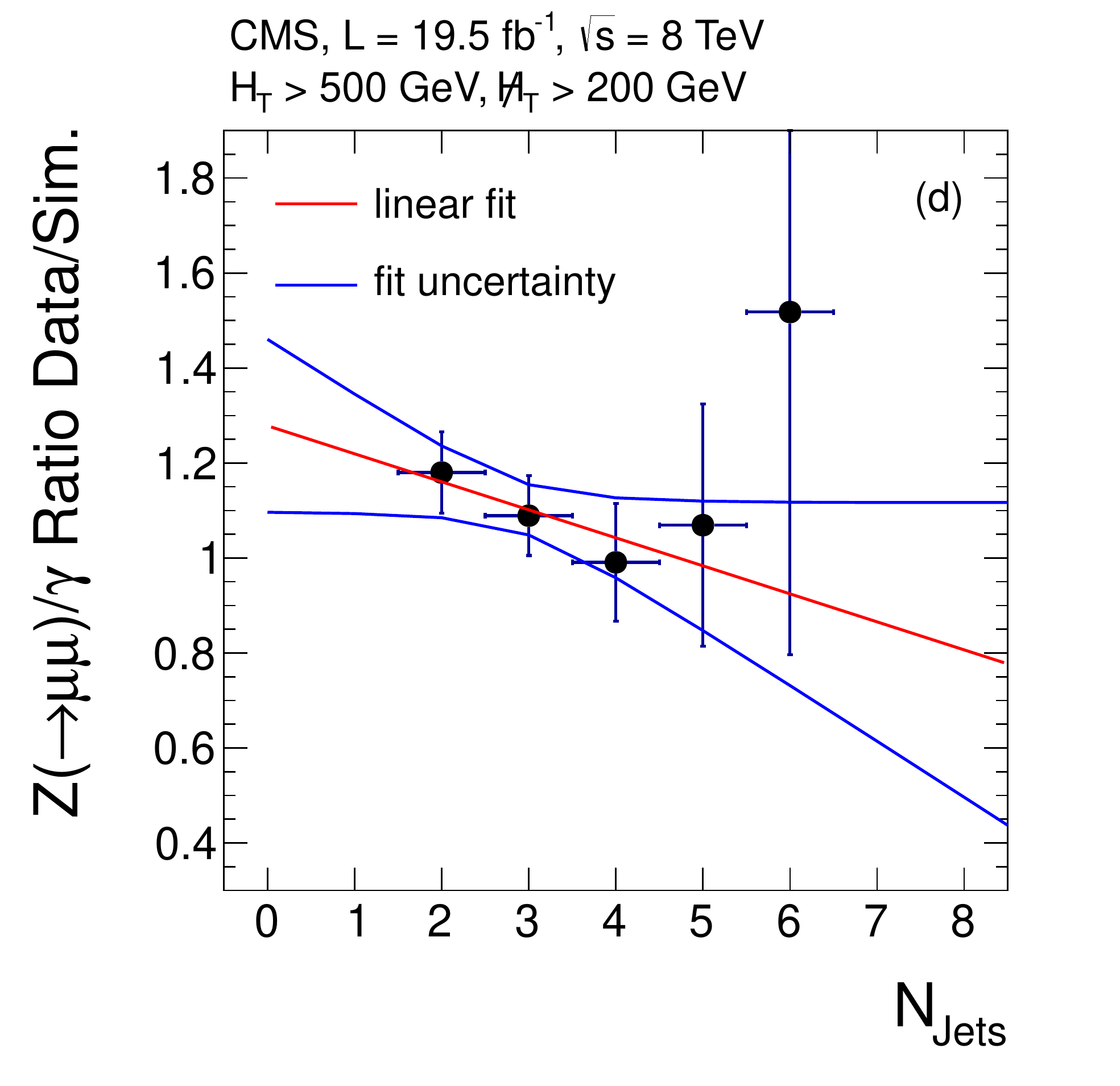}
   \caption{The simulated ratio \RZgamma as a function of (a)~\MHT, (b)~\HT,
(c)~\njets, where the values for
three \MHT bins are shown with linear fits, and
(d)~the double ratio of \RZmumugamma, using events from data
to those from simulation;
the linear fit and its uncertainty band are overlaid.}
  \label{fig:phenoratiozgamma}
\end{figure}

To mimic the missing momentum due to the neutrinos from the decay
of the \cPZ\ boson, the photon candidate is not included in the calculation of \HT and \MHT
for the $\gamma$+jets events.
The number of \znunubr{}+jets events is then estimated by correcting the number of $\gamma$+jets events
for photon acceptance and reconstruction efficiency, and scaling the result with
the ratio relating the production cross section of the two processes (\RZgamma)
in the various search regions. Therefore,
the ratio \RZgamma, which  we derive from simulation,
is studied as a function of \HT, \MHT, and \njets using events generated with
\MADGRAPH (up to four partons) that are processed through the \PYTHIA parton shower
algorithm to generate additional jets.
The ratio exhibits a strong dependence on \MHT for values
below around 500\GeV (Fig.~\ref{fig:phenoratiozgamma}(a)),
but changes by only ($12 \pm 5$)\% as
\HT varies between 500 and 1500\GeV (Fig.~\ref{fig:phenoratiozgamma}(b)),
which is the region of interest to this search.
The ratio is parametrized as a linear function of \njets in
several \MHT ranges, $200<\MHT<300$\GeV, $300<\MHT<450$\GeV, and $\MHT>450$\GeV,
as shown in Fig.~\ref{fig:phenoratiozgamma}(c). The predicted numbers
of \znunubr+jets events and uncertainties for various search regions are summarized in Table~\ref{tab:FinalEventYields}.

The theoretical uncertainty associated with \RZgamma is
estimated using \zmumubr+jets events selected from data and
simulation, by requiring two
opposite-sign muons to satisfy the muon selection and to form
an invariant mass within $\pm$20\GeV of the \cPZ\ boson mass.
The double ratio of \RZmumugamma using events from data to those
from simulation
is parametrized as a function of \njets using a linear function, as
shown in Fig.~\ref{fig:phenoratiozgamma}(d),
and is used to correct  \RZgamma
for a given jet multiplicity.
The fitting procedure results in uncertainties of 20\%, 25\%, and 45\% for the
background predicted in the search regions with \njets = 3--5, 6--7, and $\geq$8, respectively.
The difference in the modeling of photon identification and isolation in the simulation
and data leads to uncertainties of 2--5\%, 10--20\%, and 20--25\% on
the estimated number of \znunubr{}+jets events
for the three jet multiplicity intervals, respectively.
The subtraction
of events with non-prompt photons from QCD multijet events amounts to less than
a 5\% uncertainty for the final background prediction.
\subsection{Estimation of the lost-lepton background}
\label{sec:wtop_lostlepton}

The lost-lepton background is estimated from a
$\Pgm$+jets control sample, selected with the same criteria as used for the
search, except that events are required to have exactly one well-reconstructed and isolated $\Pgm$
with $\pt^{\mu}>$10\GeV. The events are collected with the same
trigger that is used to search for the signal.
The transverse mass
$\mt=\sqrt{\smash[b]{2\pt^{\mu}\ETslash[1-\cos(\Delta\phi)]}}$ is required to be
less than 100\GeV in order to select events containing $\PW\to\mu\nu$ decays
as well as to reject possible signal events. Here
$\Delta\phi$ is the azimuthal angle between the $\vec{\pt}^\mu$ and
the $\vec{\ETslash}$ directions.

Using the reconstruction and isolation efficiencies
$\epsilon_\text{reco}^{\Pe,\mu}$ and $\epsilon_\text{iso}^{\Pe,\mu}$
of the electrons and muons, the events in the isolated muon control sample
are weighted by
$\left(1/\epsilon_\text{iso}^{\mu}\right)\times
[(1-\epsilon_\text{reco}^{\Pe,\mu})/\epsilon_\text{reco}^{\mu}]$
in order to estimate the number of events with unidentified leptons, and by
$\left(\epsilon_\text{reco}^{\Pe,\mu}/\epsilon_\text{reco}^{\mu}\right)\times
[(1-\epsilon_\text{iso}^{\Pe,\mu})/\epsilon_\text{iso}^{\mu}]$
to estimate the number of events with non-isolated leptons in the signal region. The predicted
number of lost-lepton events is corrected to account for the detector and kinematic
acceptance of the muons.
The lepton efficiencies and kinematic
acceptance factors are obtained from the MC simulation of \PW+jets and \ttbar events and
are determined in bins of \njets, \HT, and \MHT.

\begin{figure}[tbh]
  \centering
    \includegraphics[width=0.32\textwidth]{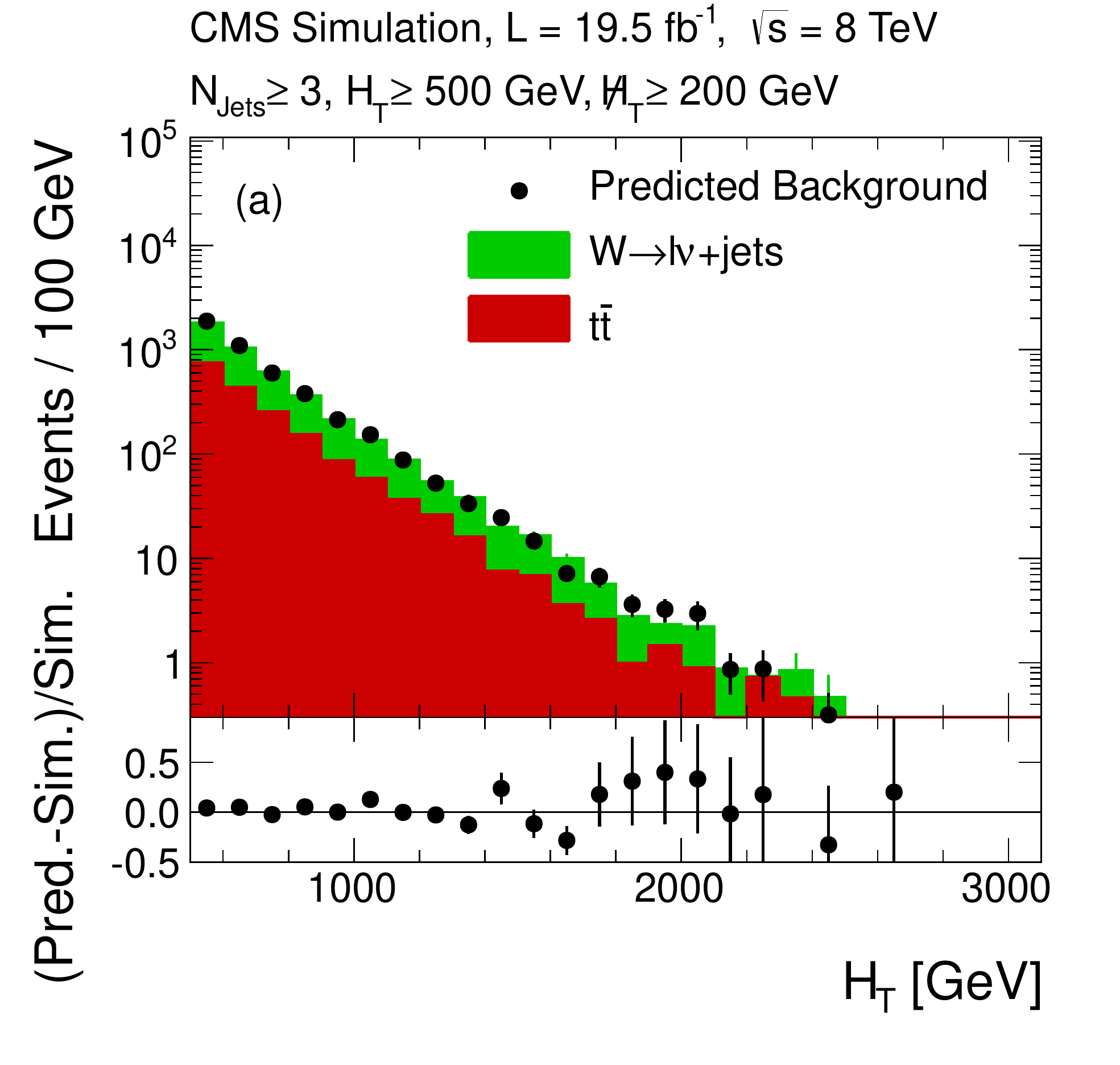}
    \includegraphics[width=0.32\textwidth]{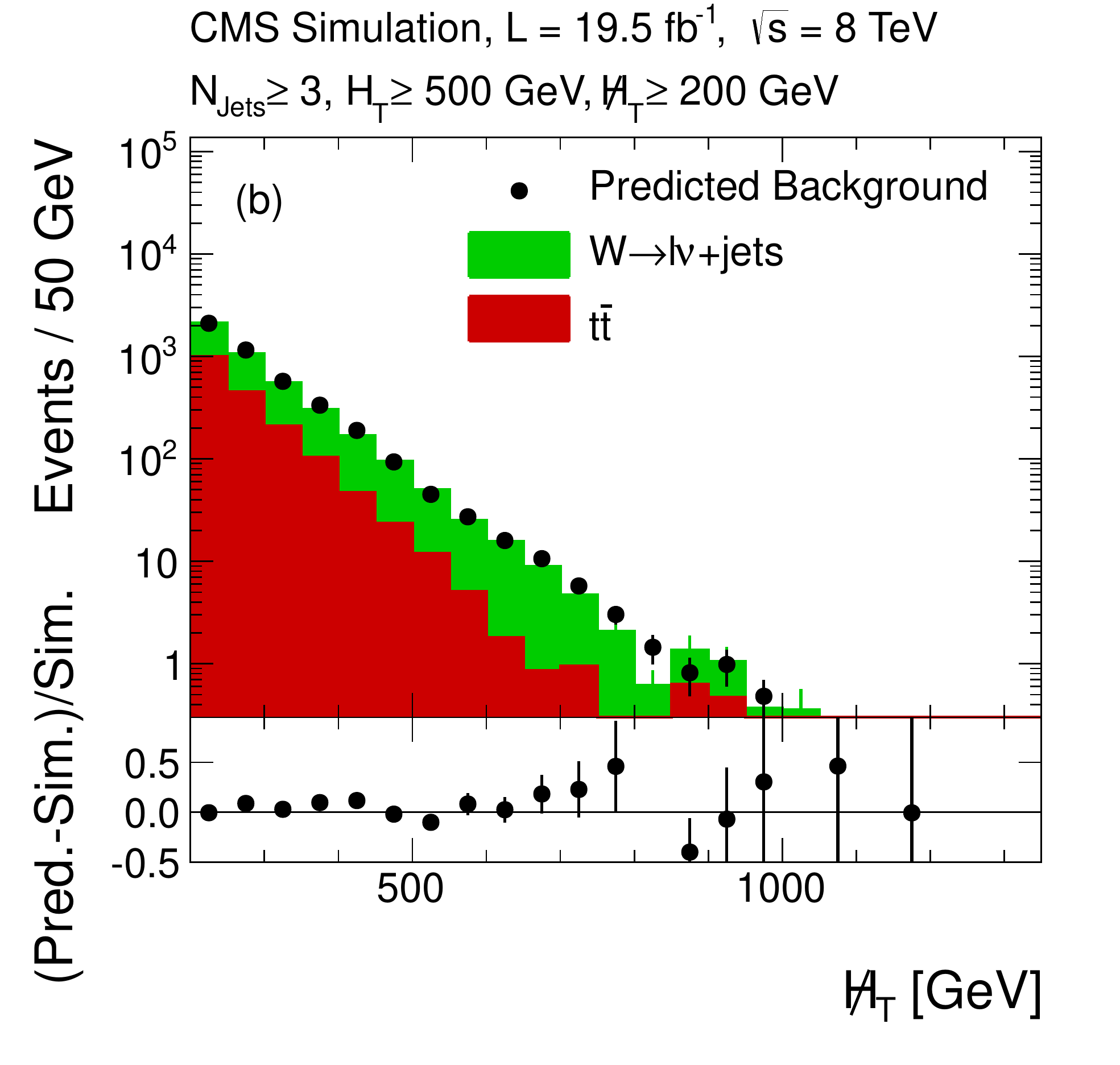}
    \includegraphics[width=0.32\textwidth]{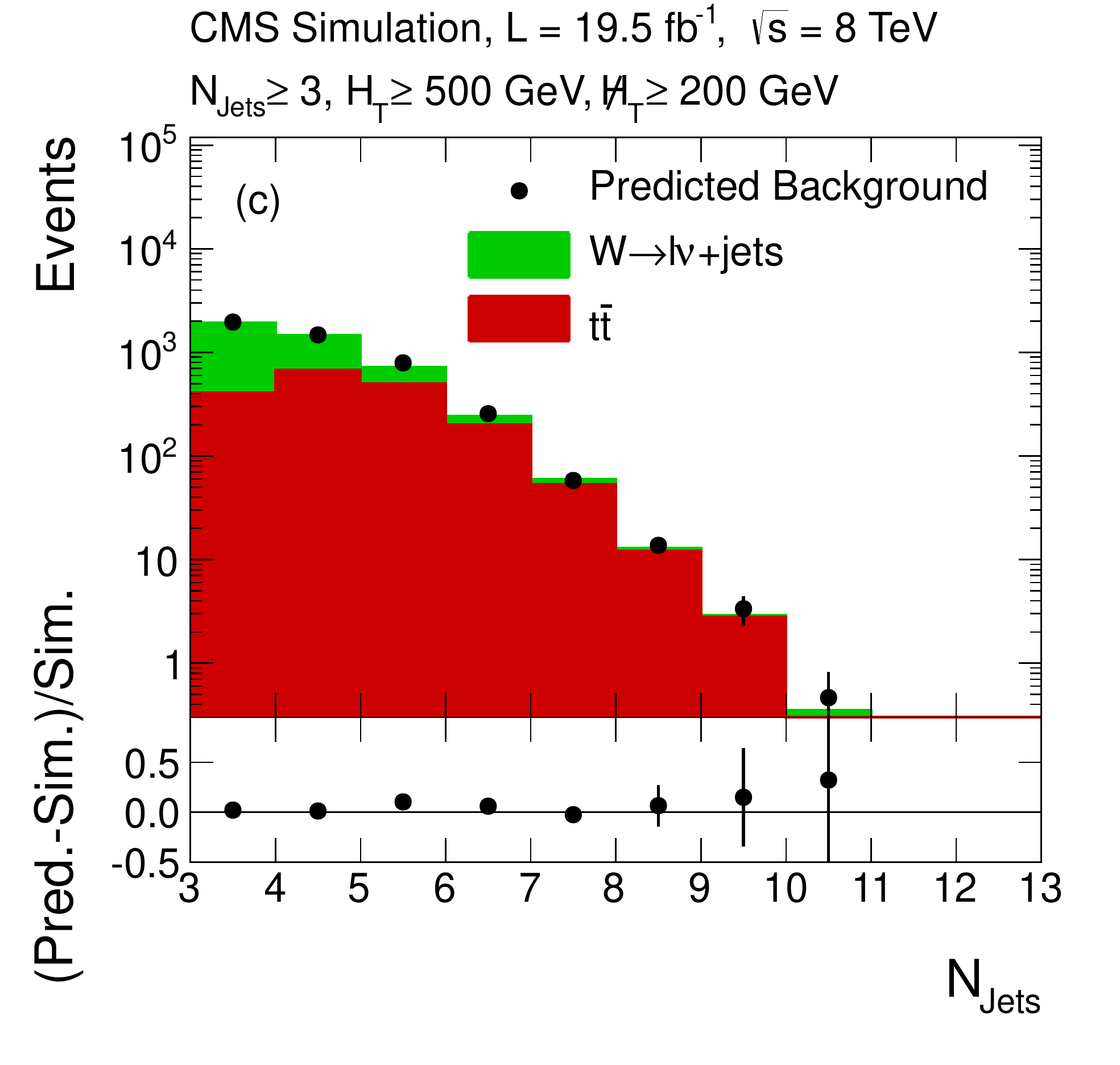}
    \caption{Predicted (a)~\HT, (b)~\MHT, and (c)~\njets distributions found from
     applying the lost-lepton background evaluation method to simulated
     \ttbar and \wpj events (solid points) in comparison to the genuine \ttbar and \wpj
     background from simulation (shaded curves).
     Only statistical uncertainties are shown.}
    \label{fig:ClosureTestMC}
\end{figure}

This method is validated using simulated \ttbar and \PW+jets events.
The single-muon events selected from the simulated samples are used to predict
the number of background events expected in the zero-lepton search regions. The resulting \HT, \MHT, and \njets distributions are compared in Fig.~\ref{fig:ClosureTestMC}
to the genuine ones obtained from \ttbar and \PW+jets
events simulated at the detector level.  The predicted
distributions closely resemble the genuine ones.

The number of lost-lepton events predicted from data using the method
described above, and the corresponding uncertainties,
are listed in Table~\ref{tab:FinalEventYields} for each search region.
The dominant uncertainties arise from the
limited number of single-muon events in
most of the search regions.
The differences in lepton reconstruction and isolation
efficiencies between data and MC simulation are evaluated
using a ``tag-and-probe'' method~\cite{EWK-10-005}
on  \zmumubr{}+jets events.
The lepton reconstruction and isolation efficiencies are measured
in bins of lepton \pt and $\Delta R$
relative to the closest jet. This method renders these efficiencies insensitive
to the kinematic differences between \zellellbr{}+jets events
and \ttbar and \PW+jets events.
Relative differences between the predictions
using efficiencies extracted from data and MC simulation result in
10--25\%, 10--30\%, and 15--24\% uncertainties for the predicted background
for various \HT and \MHT search bins with \njets = 3--5, 6--7, and $\geq$8, respectively.
An additional uncertainty of 15\% for \njets = 3--5 and 40\% for $\njets\geq6$
is assigned based on the statistical precision of the validation of this background estimation method.
Variation of the PDFs following the procedure of Ref.~\cite{Botje:2011sn}
affects the muon acceptance, and leads to an uncertainty of less than 4\% on the final prediction.
Any mismodeling of anomalous \ETslash~\cite{METJINST} affects the simulated $\mt$
and results in 3\% uncertainty for the predicted lost-lepton background.

\subsection{Estimation of the hadronic \texorpdfstring{$\tau$}{tau} lepton background}
\label{sec:wtop_hadronictau}

The \tauh background is estimated
from a sample of $\Pgm$+jets events, selected with an inclusive single $\Pgm$ or
$\mu+{\geq}2$-jet trigger, by requiring exactly one $\Pgm$ with $\pt>20\GeV$
and $\abs{\eta}<2.1$. As in the estimation of the lost-lepton background,
only events with $\mt<100$\GeV are considered.
The $\mu$+jets and \tauh+jets events arise from the same physics processes; hence
the hadronic component of the two samples is the same aside from the response of the
detector to a muon or a $\tauh$ jet.
To account for this difference, the muon is replaced by a simulated \tauh jet,
whose \pt value is randomly sampled from an MC response function, $\pt^{\text{Jet}}/\pt^\tau$.
Here, the $\pt^\tau$ is the transverse momentum of a generated hadronically decaying $\tau$ lepton
selected from simulated $\ttbar$ and $\wtaunubr$+jets events and $\pt^{\text{Jet}}$
is that of a reconstructed jet matching the $\tau$ lepton in $\eta$--$\phi$ space.
In order to sample the response function completely, this procedure is repeated
one hundred times for each event.
The \njets, \HT, and \MHT values of the events are recalculated, now including this $\tauh$ jet,
and search region selection criteria are applied to predict the \tauh background.
The predicted background is corrected for the trigger efficiency, muon selection efficiency,
kinematic and detector acceptance, and
the ratio of branching fractions
$\mathcal{B}(\PW\to\tauh\nu)/\mathcal{B}(\PW\to\mu\nu)= 0.6476 \pm 0.0024$~\cite{PDG}.
The muon isolation and reconstruction efficiencies
are obtained from MC simulation of \PW+jets and \ttbar events
in bins of lepton \pt and $\Delta R$
relative to the closest jet. To account for
the difference in
efficiencies measured in data and MC simulation, the predicted numbers of $\tauh$+jets events are
corrected by 4.9\%, 4.7\%, and 3.5\% for \njets=3--5, 6--7, and $\geq$8, respectively.
The predicted $\tauh$ background and uncertainties are shown in Table~\ref{tab:FinalEventYields}
for all the search regions.

\begin{figure}[htb]
  \centering
    \includegraphics[width=0.32\textwidth] {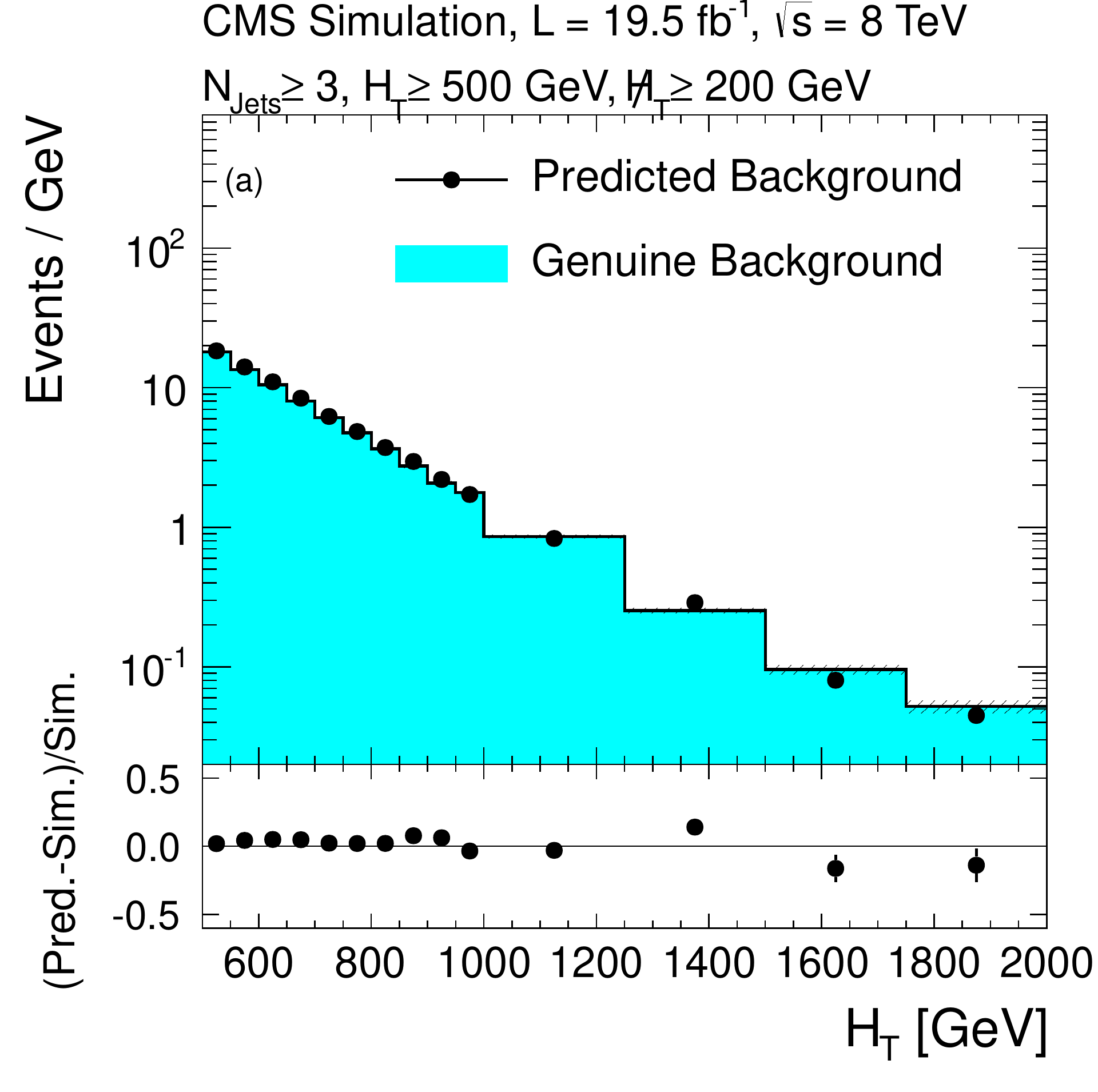} \includegraphics[width=0.32\textwidth] {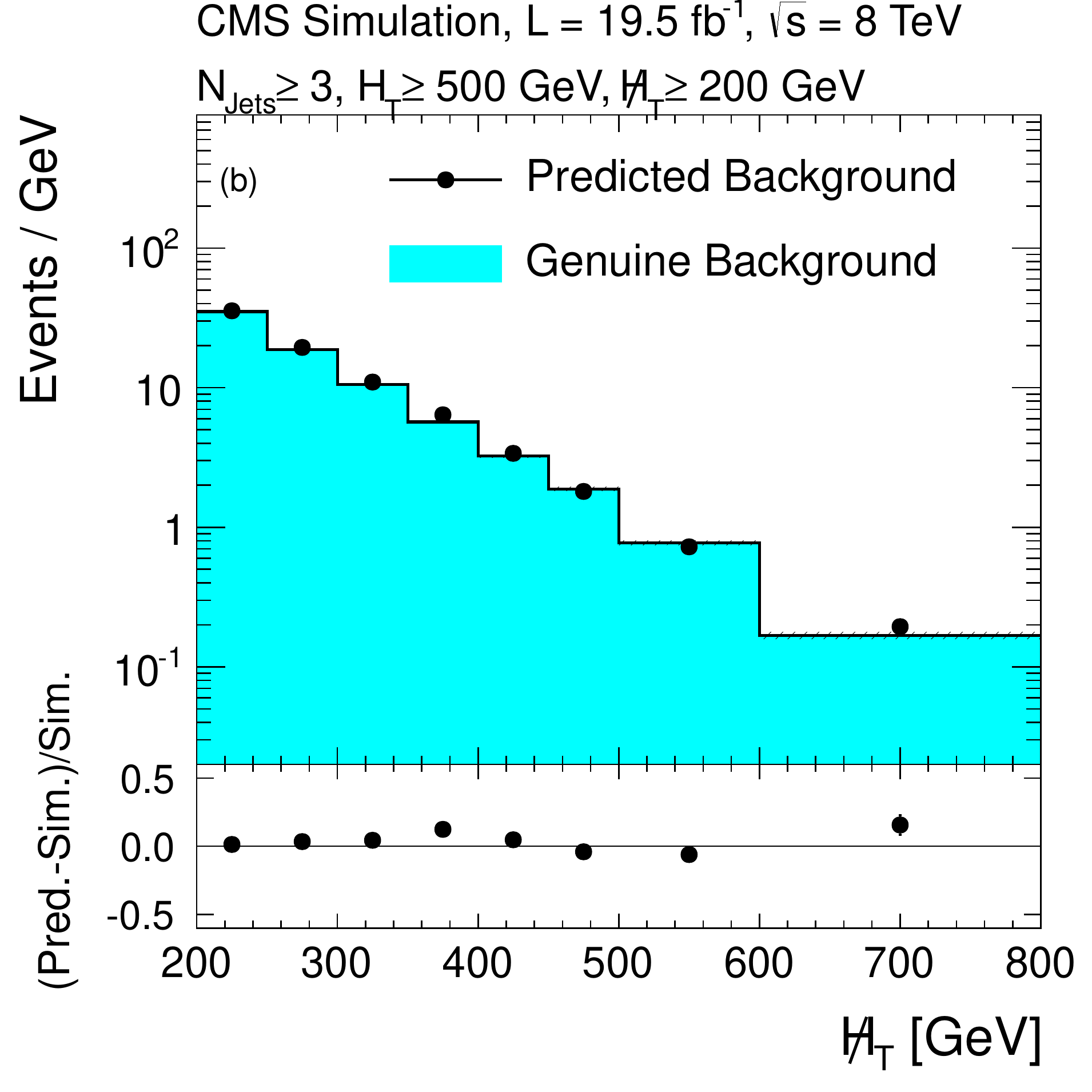}
    \includegraphics[width=0.32\textwidth] {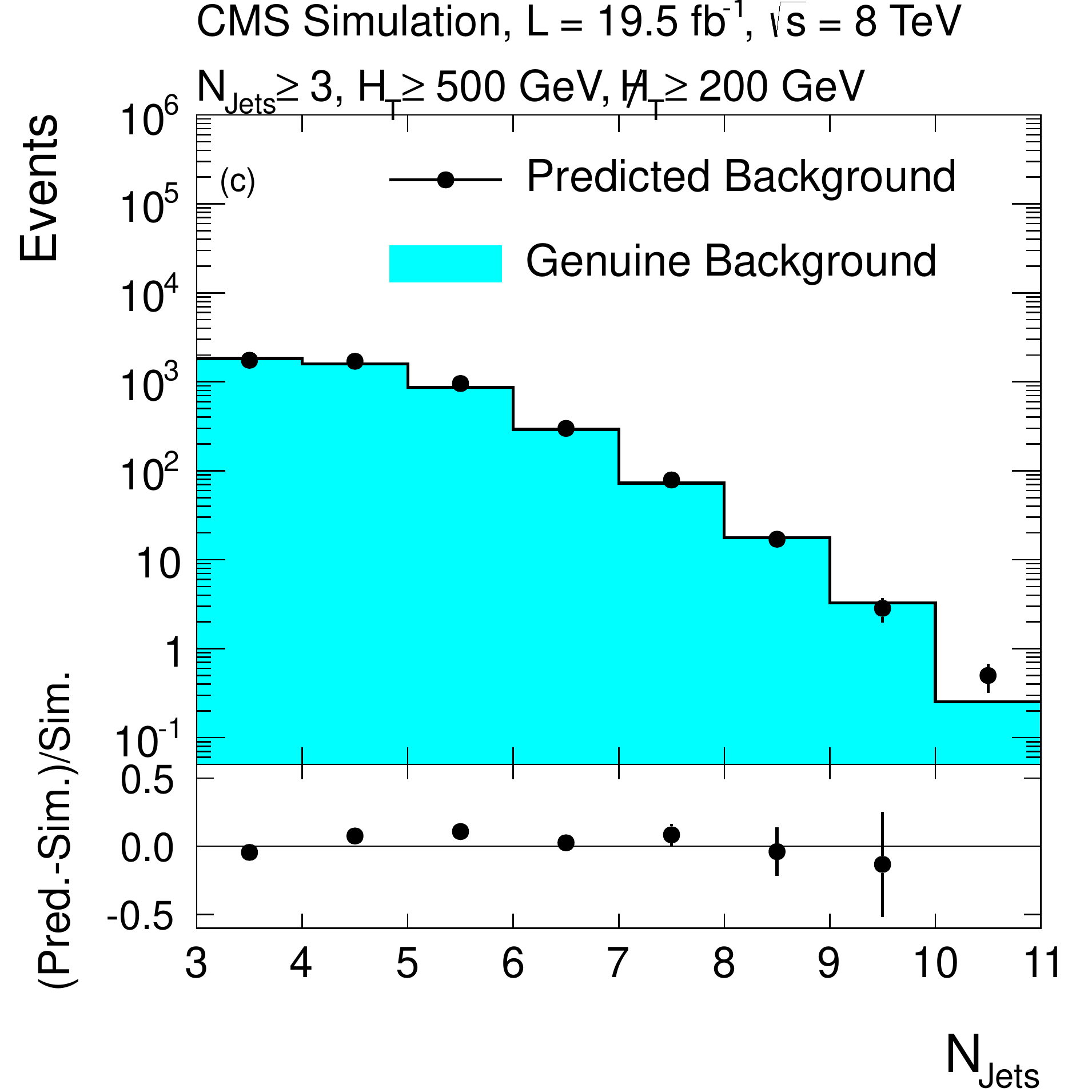}
    \caption{Predicted (a)~\HT, (b)~\MHT, and (c)~\njets distributions found from applying
    the $\tauh$ background evaluation method to simulated \ttbar and \wpj events
    (solid points) in comparison to the genuine \ttbar and \wpj background from
    simulation (shaded curve).
    Only statistical uncertainties are shown.}
    \label{fig:TAUpredictionMC}
\end{figure}

The \tauh background estimation method is validated by applying it to simulated \PW+jets
and \ttbar MC samples. The results are shown in
Fig.~\ref{fig:TAUpredictionMC} in comparison to the genuine
\tauh background from the simulated events.
To evaluate the performance of the method for events with varying hadronic activity,
the method is validated in each search bin.
Uncertainties of 10\%, 20\%, and 20\% are assigned to the predicted rates
for events with \njets=3--5, 6--7, and $\geq$8 respectively, mainly
to reflect the level of statistical precision for this validation.
Due to the multiple sampling of the response template, the statistical uncertainty of the
prediction is evaluated with a set of pseudo-experiments using a bootstrap technique~\cite{tEFR82a}.
Relative differences between the predictions
using efficiencies extracted from data and MC result in 2--20\% uncertainties
across the various search bins.
Other systematic uncertainties
arise from the geometrical and kinematic acceptance for the muons (3\%), and
the $\tau$-jet response function (1--15\%).
An uncertainty of 1--8\% is assigned to
account for possible differences between data and MC simulation for the acceptance of
the $\mt$ selection.

\subsection{Estimation of the QCD multijet background}
\label{sec:qcd}

The background from QCD multijet events is evaluated with the
``rebalance and smear'' method ~\cite{RA2,RA2_2011}, using data
samples recorded with \HT thresholds ranging from 350 to 650\GeV.
The events, recorded with a trigger prescaled by a factor $k$,
are sampled $k$ times to create seed events as described below.

In the rebalance step, the momenta of the jets with
$\pt>10$\GeVc in each event are adjusted within the jet-\pt-resolution values,
using a kinematic fit, such that the events are balanced in the transverse plane.
Considering only jets with \pt above a certain threshold introduces an additional
imbalance in the event, which results in larger \pt for the rebalanced jets
than the expected true value. This effect is compensated by
scaling the rebalanced jets by a \pt-dependent factor derived
by comparing rebalanced and generator-level jets in the simulation.
The scaling factors derived using either
\PYTHIA or \MADGRAPH, and with different average
pileup interactions, are found to be similar.
The jets in the rebalanced events are then smeared using jet \pt response
functions, which are obtained from MC simulation as a function of \pt and $\eta$,
and adjusted to match those determined from
dijet and $\gamma$+jets data~\cite{JETJINST}. The  QCD multijet background is
predicted by applying selection
criteria on the kinematic quantities calculated from the smeared jets.
The procedure is
repeated one hundred times to evaluate the average prediction and
its statistical uncertainty in each search region.

\begin{figure}[htb]
  \begin{center}
    \includegraphics[width=0.32\textwidth]{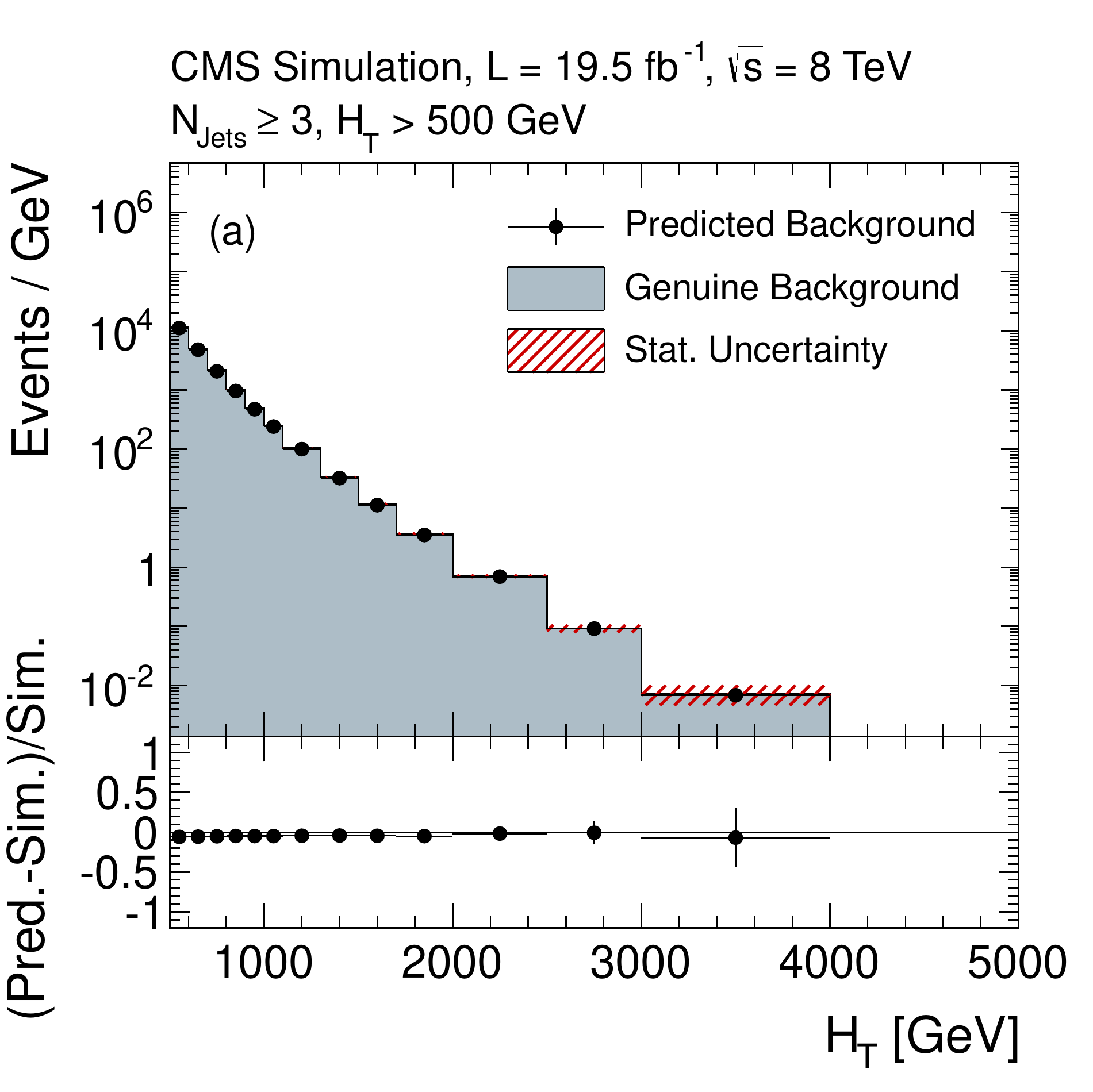}
    \includegraphics[width=0.32\textwidth]{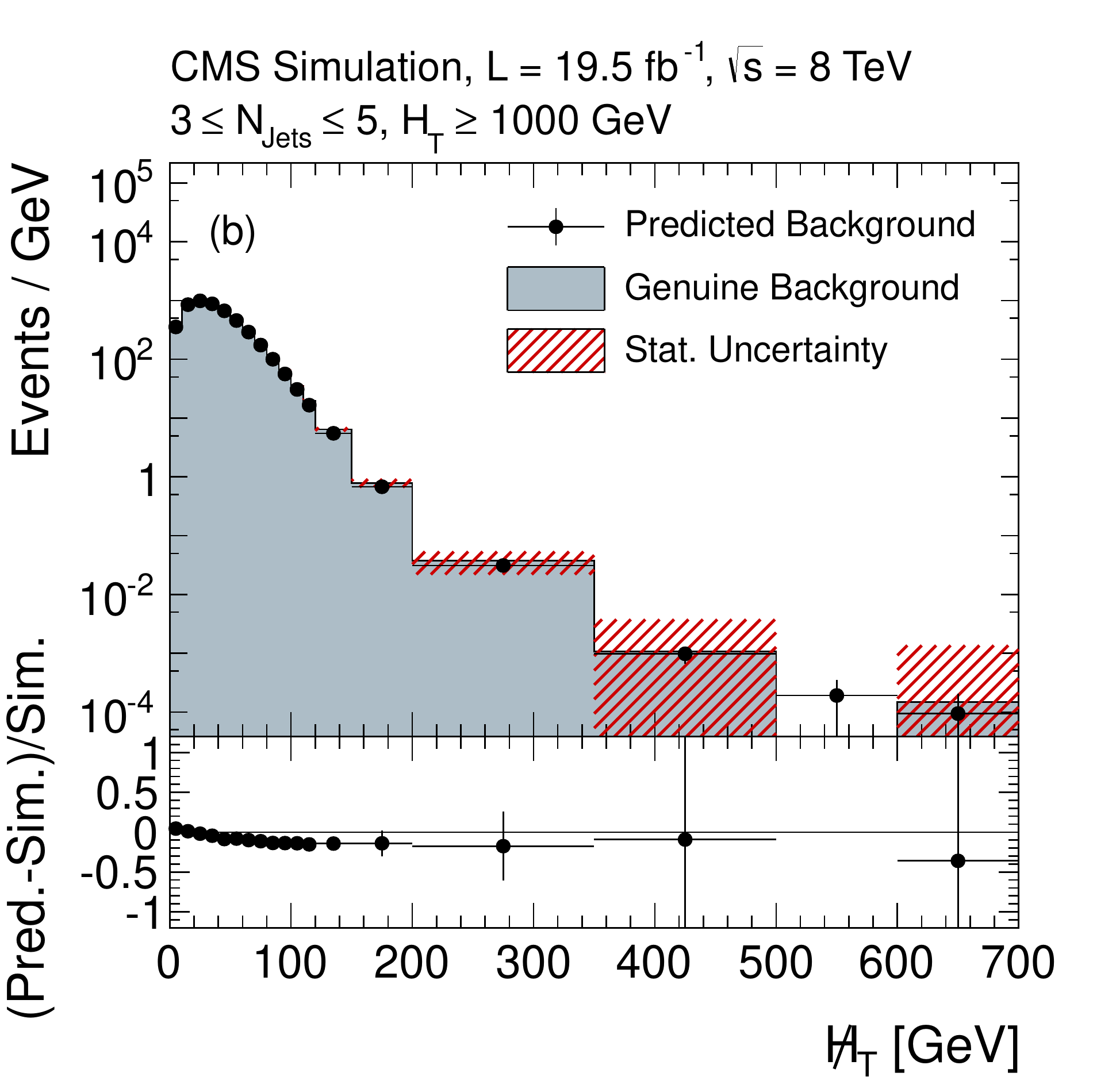}
    \includegraphics[width=0.32\textwidth]{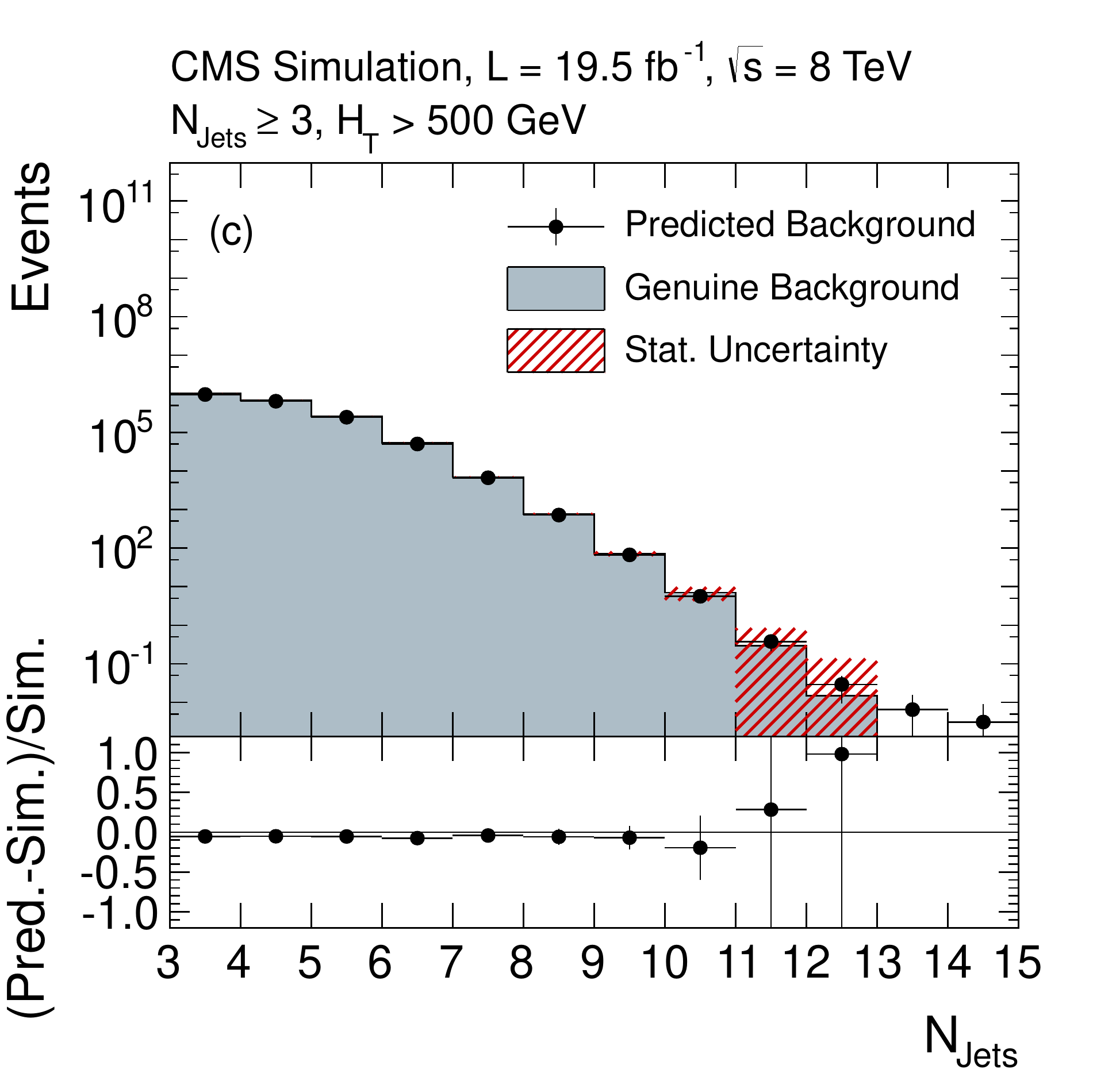}
    \caption{Predicted (a)~\HT, (b)~\MHT, and (c)~\njets distributions found from applying the ``rebalance-and-smear" method to simulated QCD multijet events (solid points) in comparison with the genuine QCD multijet background from simulation (shaded curve). The distributions are shown for events that satisfy the baseline selection, except that the \MHT selection is not applied, and in addition $\HT>1000\GeV$ is required for the events used in the \MHT distribution. The statistical uncertainties are indicated by the hatched band for the
  expectation and by error bars for the prediction.}
    \label{fig:QCDClosureTestMC}
  \end{center}
\end{figure}

The method is validated using simulated QCD multijet events.
Comparisons of the \HT, \MHT, and \njets distributions from the MC
simulation to those predicted by the rebalance-and-smear
method on the same simulated events are shown in Fig.~\ref{fig:QCDClosureTestMC}.
A systematic uncertainty of 11--86\% is assigned based on the
statistical precision attributed to the validation procedure,
which is performed both in the search regions and in QCD-enriched data
control regions defined either by $100 <\MHT<200$\GeV or by inverting
the $\abs{\Delta\phi(\ptvec^{\text{Jet1,2,3}}, \vec{\MHT})}$ selection.
Due to the limited number of events in individual search bins, this uncertainty
is evaluated for each jet multiplicity bin for \HT smaller or larger than 1000\GeV, inclusive over \MHT.
The uncertainty due to differences in the core and tails of the jet
response functions between data and simulation results in uncertainties
of 10--30\% and 20--35\%, respectively.
An uncertainty of 3\%, 8\%, and 35\% is assigned for search regions with
\njets = 3--5, 6--7, and $\geq$8, respectively, to account for the effect
of pileup.
The predicted QCD multijet background contributions to the search bins
along with associated uncertainties are given in Table~\ref{tab:FinalEventYields}.

\section{Results and interpretation}
\label{sec:sensitivity}

\begin{figure}[ptbh]
\centering
\includegraphics[width=\textwidth]{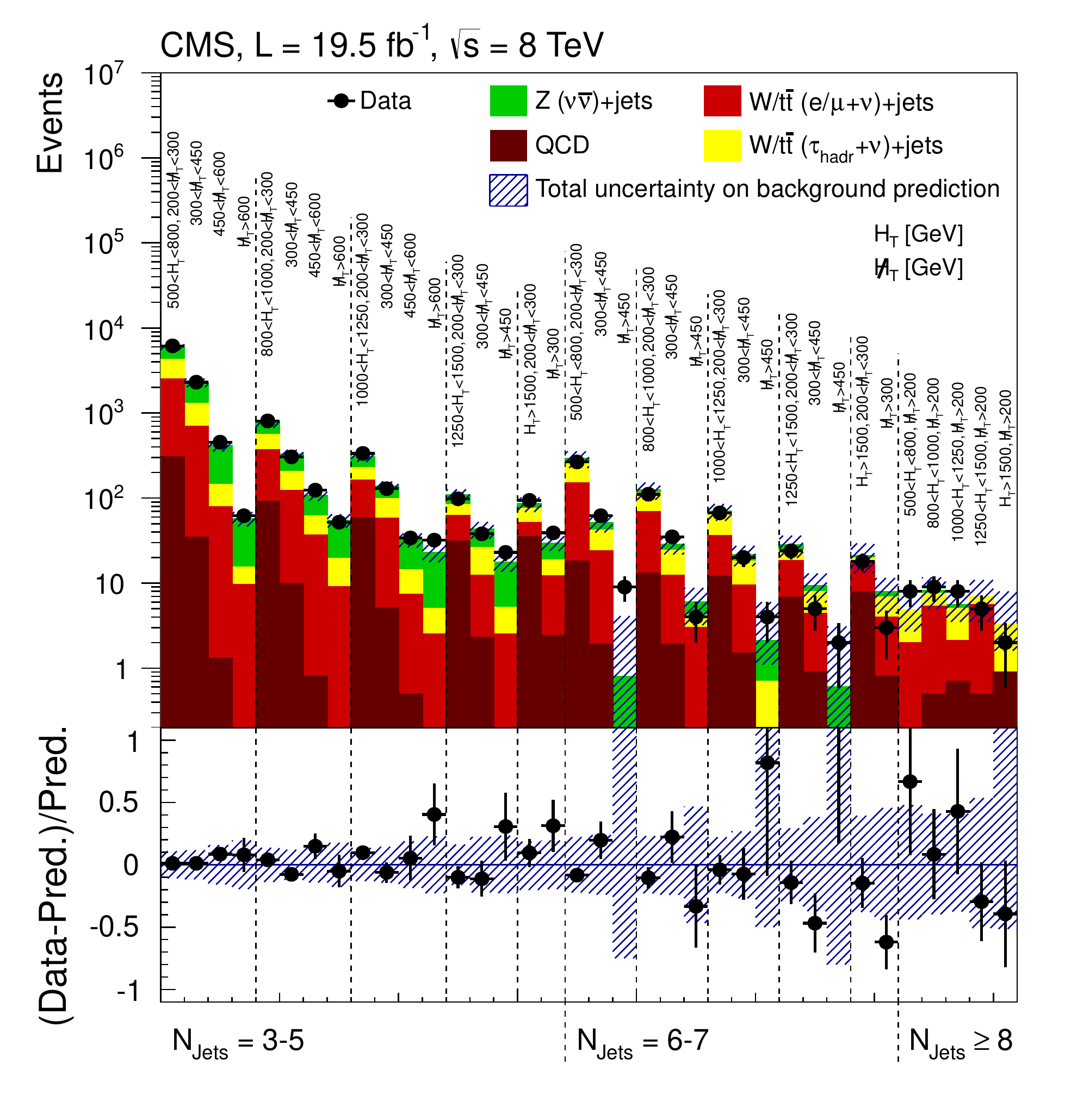}
\caption{Summary of the observed number of events in each of the 36 search regions in comparison to the corresponding background prediction. The hatched region shows the total uncertainty of the background prediction.}
\label{fig:htmhtcombined}
\end{figure}

The predicted background event yields and the number of observed events
are summarized in Table~\ref{tab:FinalEventYields} and Fig.~\ref{fig:htmhtcombined}
for the 36 search regions. The data are consistent
with the expected background contributions from SM processes. A slight excess of events
is observed in the search bin with $\njets = 6$--7, $\HT = 500$--800\GeV, and $\MHT> 450$\GeV,
which is insignificant when the probability to observe a statistical fluctuation
as large or larger in any of the search regions is considered.

\begin{table}[p]
\centering
\topcaption{Predicted event yields for the different background components in the search regions
defined by \HT, \MHT and \njets. The uncertainties of the different
background sources are added in quadrature to obtain the total uncertainties.
}
\label{tab:FinalEventYields}
\renewcommand{\arraystretch}{1.2}
\resizebox{\textwidth}{!}
{
\begin{tabular}{lcc|r@{}c@{}l|r@{}c@{}l|r@{}c@{}l|r@{}c@{}l|r@{}c@{}l|r}
              \multicolumn{3}{c|}{Selection}
            & \multicolumn{3}{c|}{$\znunu$}
            & \multicolumn{3}{c|}{$\ttbar/\PW$}
            & \multicolumn{3}{c|}{$\ttbar/\PW$}
	    & \multicolumn{3}{c|}{QCD}
            & \multicolumn{3}{c|}{Total }
	    & \multicolumn{1}{c}{Data}          \\

            \njets & \HT [\GeVns{}] &\MHT [\GeVns{}]
            & \multicolumn{3}{c|}{}
	    & \multicolumn{3}{c|}{$\to \Pe,\mu+$X}
            & \multicolumn{3}{c|}{$\to \tauh+$X}
	    & \multicolumn{3}{c|}{}
            & \multicolumn{3}{c|}{background}
	    & \multicolumn{1}{c}{}  \\
\hline
3--5     & 500--800    & 200--300  & 1820   &$\pm$& 390         & 2210   &$\pm$& 450        &1750   &$\pm$& 210        &310   &$\pm$& 220         & 6090   &$\pm$& 670        &  6159  \\
3--5     & 500--800    & 300--450  &  990   &$\pm$& 220         &  660   &$\pm$& 130        & 590   &$\pm$&  70        & 40   &$\pm$&  20         & 2280   &$\pm$& 270        &  2305  \\
3--5     & 500--800    & 450--600  &  273   &$\pm$&  63         &   77   &$\pm$&  17        &  66.3 &$\pm$&   9.5      &  1.3 & \multicolumn{2}{@{}l|}{$^{+1.5}_{-1.3}$} & 418    &$\pm$&  66        &   454  \\
3--5     & 500--800    & $>$600   &   42   &$\pm$&  10         &   9.5 &$\pm$&   4.0      &   5.7 &$\pm$&   1.3      &  0.1 & \multicolumn{2}{@{}l|}{$^{+0.3}_{-0.1}$} & 57.4  &$\pm$&  11.2  &    62  \\ \hline
3--5     & 800--1000   & 200--300  &  216   &$\pm$&  46         &  278   &$\pm$&  62        & 192   &$\pm$&  33        & 92   &$\pm$&  66         & 777   &$\pm$& 107         &   808  \\
3--5     & 800--1000   & 300--450  &  124   &$\pm$&  26         &  113   &$\pm$&  27        &  84   &$\pm$&  12        &  9.9 &$\pm$&   7.4       & 330   &$\pm$&  40         &   305  \\
3--5     & 800--1000   & 450--600  &   47   &$\pm$&  11         &   36.1 &$\pm$&   9.9      &  24.1 &$\pm$&   3.6      &  0.8 & \multicolumn{2}{@{}l|}{$^{+1.3}_{-0.8}$} & 108   &$\pm$&  15         &   124  \\
3--5     & 800--1000   & $>$600   &   35.3 &$\pm$&   8.8       &    9.0 &$\pm$&   3.7      &  10.3 &$\pm$&   2.0      &  0.1 & \multicolumn{2}{@{}l|}{$^{+0.4}_{-0.1}$} & 54.8 &$\pm$&    9.7       &    52  \\ \hline
3--5     & 1000--1250  & 200--300  &   76   &$\pm$&  17        &  104   &$\pm$&  26        &  66.5 &$\pm$&   9.9      & 59   &$\pm$&  25         & 305   &$\pm$&  41         &   335  \\
3--5     & 1000--1250  & 300--450  &   39.3 &$\pm$&   8.9       &   52   &$\pm$&  14        &  41   &$\pm$&  11        &  5.1 &$\pm$&   2.7       & 137   &$\pm$&  20         &   129  \\
3--5     & 1000--1250  & 450--600  &   18.1 &$\pm$&   4.7       &    6.9 &$\pm$&   3.2      &   6.8 &$\pm$&   2.0      &  0.5 & \multicolumn{2}{@{}l|}{$^{+0.7}_{-0.5}$} & 32.3 &$\pm$&   6.1        &    34  \\
3--5     & 1000--1250  & $>$600   &   17.8 &$\pm$&   4.8       &    2.4 &$\pm$&   1.8      &   2.5 &$\pm$&   0.8      &  0.1 & \multicolumn{2}{@{}l|}{$^{+0.3}_{-0.1}$} & 22.8 &$\pm$&   5.2        &    32  \\ \hline
3--5     & 1250--1500  & 200--300  &   25.3 &$\pm$&   6.0       &   31.0 &$\pm$&   9.5      &  21.3 &$\pm$&   4.1      & 31   &$\pm$&  13         & 109   &$\pm$&  18         &    98  \\
3--5     & 1250--1500  & 300--450  &   16.7 &$\pm$&   4.3       &   10.1 &$\pm$&   4.4      &  13.7 &$\pm$&   7.1      &  2.3 &$\pm$&   1.6       & 42.8 &$\pm$&   9.5        &    38  \\
3--5     & 1250--1500  & $>$450   &   12.3 &$\pm$&   3.5       &    2.3 &$\pm$&   1.7      &   2.7 &$\pm$&   1.2      &  0.2 & \multicolumn{2}{@{}l|}{$^{+0.5}_{-0.2}$} & 17.6 &$\pm$&   4.1        &    23  \\ \hline
3--5     & $>$1500    & 200--300  &   10.5 &$\pm$&   2.9       &   16.7 &$\pm$&   6.2      &  23.5 &$\pm$&   5.6      & 35   &$\pm$&  14         & 86   &$\pm$&  17          &    94  \\
3--5     & $>$1500    & $>$300   &   10.9 &$\pm$&   3.1       &    9.7 &$\pm$&   4.3      &   6.6 &$\pm$&   1.4      &  2.4 &$\pm$&   2.0       & 29.7 &$\pm$&   5.8        &    39  \\ \hline \hline
6--7     & 500--800    & 200--300  &   22.7 &$\pm$&   6.4       &  133   &$\pm$&  59        & 117   &$\pm$&  25        & 18.2 &$\pm$&   9.2       & 290   &$\pm$&  65         &   266  \\
6--7     & 500--800    & 300--450  &    9.9 &$\pm$&   3.2       &   22   &$\pm$&  11        &  18.0 &$\pm$&   5.1      &  1.9 &$\pm$&   1.7       & 52   &$\pm$&  12          &    62  \\
6--7     & 500--800    & $>$450   &    0.7 &$\pm$&   0.6       &    0.0 & \multicolumn{2}{@{}l|}{$^{+3.2}_{-0.0}$}&   0.1 & \multicolumn{2}{@{}l|}{$^{+0.5}_{-0.1}$}&  0.0 & \multicolumn{2}{@{}l|}{$^{+0.1}_{-0.0}$} & 0.8 &\multicolumn{2}{@{}l|}{$^{+3.3}_{-0.6}$}    &     9  \\ \hline
6--7     & 800--1000   & 200--300  &    9.1 &$\pm$&   3.0       &   56   &$\pm$&  25        &  46   &$\pm$&  11        & 13.1 &$\pm$&   6.6       & 124   &$\pm$&  29       &   111  \\
6--7     & 800--1000   & 300--450  &    4.2 &$\pm$&   1.7       &   10.4 &$\pm$&   5.5      &  12.0 &$\pm$&   3.6      &  1.9 &$\pm$&   1.4       & 28.6 &$\pm$&   6.9        &    35  \\
6--7     & 800--1000   & $>$450   &    1.8 &$\pm$&   1.0       &    2.9 &$\pm$&   2.5      &   1.2 &$\pm$&   0.8      &  0.1 & \multicolumn{2}{@{}l|}{$^{+0.4}_{-0.1}$} & 6.0 &$\pm$&   2.8         &     4  \\ \hline
6--7     & 1000--1250  & 200--300  &    4.4 &$\pm$&   1.7       &   24   &$\pm$&  12        &  29.5 &$\pm$&   7.8      & 11.9 &$\pm$&   6.0  &  70  &$\pm$&  16               &    67  \\
6--7     & 1000--1250  & 300--450  &    3.5 &$\pm$&   1.5       &    8.0 &$\pm$&   4.7      &   8.6 &$\pm$&   2.7      &  1.5 &$\pm$&   1.5  & 21.6 &$\pm$&   5.8             &    20  \\
6--7     & 1000--1250  & $>$450   &    1.4 &$\pm$&   0.8       &    0.0 & \multicolumn{2}{@{}l|}{$^{+3.6}_{-0.0}$}&   0.6 & \multicolumn{2}{@{}l|}{$^{+0.8}_{-0.6}$}&  0.1 & \multicolumn{2}{@{}l|}{$^{+0.4}_{-0.1}$} & 2.2 &\multicolumn{2}{@{}l|}{$^{+3.8}_{-1.1}$}    &     4  \\ \hline
6--7     & 1250--1500  & 200--300  &    3.3 &$\pm$&   1.4       &   11.5 &$\pm$&   6.5      &   6.4 &$\pm$&   2.7      &  6.8 &$\pm$&   3.9       & 28.0 &$\pm$&   8.2        &    24  \\
6--7     & 1250--1500  & 300--450  &    1.4 &$\pm$&   0.8       &    3.5 &$\pm$&   2.6      &   3.5 &$\pm$&   1.9      &  0.9 & \multicolumn{2}{@{}l|}{$^{+1.3}_{-0.9}$} & 9.4 &$\pm$&   3.6         &     5  \\
6--7     & 1250--1500  & $>$450   &    0.4 & $\pm$&   0.4      &    0.0 & \multicolumn{2}{@{}l|}{$^{+2.5}_{-0.0}$}&   0.1 & \multicolumn{2}{@{}l|}{$^{+0.5}_{-0.1}$}&  0.1 & \multicolumn{2}{@{}l|}{$^{+0.3}_{-0.1}$} & 0.5 &\multicolumn{2}{@{}l|}{$^{+2.6}_{-0.4}$}    &     2  \\ \hline
6--7     & $>$1500    & 200--300  &    1.3 &$\pm$&   0.8       &   10.0 &$\pm$&   6.9      &   2.0 &$\pm$&   1.2      &  7.8 &$\pm$&   4.0       & 21.1 &$\pm$&   8.1        &    18  \\
6--7     & $>$1500    & $>$300   &    1.1 &$\pm$&   0.7       &    3.2 &$\pm$&   2.8      &   2.8 &$\pm$&   1.9      &  0.8 & \multicolumn{2}{@{}l|}{$^{+1.1}_{-0.8}$} & 7.9 &$\pm$&   3.6         &     3  \\ \hline \hline
$\geq$8 & 500--800    & $>$200   &    0.0 & \multicolumn{2}{@{}l|}{$^{+0.8}_{-0.0}$} &    1.9 &$\pm$&   1.5      &   2.8 &$\pm$&   1.4      &  0.1 & \multicolumn{2}{@{}l|}{$^{+0.4}_{-0.1}$} & 4.8 &\multicolumn{2}{@{}l|}{$^{+2.3}_{-2.1}$}    &     8  \\
$\geq$8 & 800--1000   & $>$200   &    0.6 &$\pm$&   0.6       &    4.8 &$\pm$&   2.9      &   2.3 &$\pm$&   1.2      &  0.5 & \multicolumn{2}{@{}l|}{$^{+0.9}_{-0.5}$} & 8.3 &\multicolumn{2}{@{}l|}{$^{+3.4}_{-3.3}$}    &     9  \\
$\geq$8 & 1000--1250  & $>$200   &    0.6 &$\pm$&   0.5       &    1.4 & \multicolumn{2}{@{}l|}{$^{+1.5}_{-1.4}$}&   2.9 &$\pm$&   1.3      &  0.7 & \multicolumn{2}{@{}l|}{$^{+1.0}_{-0.7}$} & 5.6 &\multicolumn{2}{@{}l|}{$^{+2.3}_{-2.1}$}    &     8  \\
$\geq$8 & 1250--1500  & $>$200   &    0.0 & \multicolumn{2}{@{}l|}{$^{+0.9}_{-0.0}$} &    5.1 &$\pm$&   3.5      &   1.4 &$\pm$&   0.9      &  0.5 & \multicolumn{2}{@{}l|}{$^{+0.9}_{-0.5}$} & 7.1 &\multicolumn{2}{@{}l|}{$^{+3.8}_{-3.6}$}    &     5  \\
$\geq$8 & $>$1500    & $>$200   &    0.0 & \multicolumn{2}{@{}l|}{$^{+0.7}_{-0.0}$} &    0.0 & \multicolumn{2}{@{}l|}{$^{+4.2}_{-0.0}$}&   2.4 &$\pm$&   1.4      &  0.9 & \multicolumn{2}{@{}l|}{$^{+1.3}_{-0.9}$} & 3.3 &\multicolumn{2}{@{}l|}{$^{+4.7}_{-1.7}$}    &     2  \\ \hline \end{tabular}
}
\end{table}

The results are interpreted in the context of simplified models ~\cite{Alwall:2008ag,SMSPaper} of
pair production of squarks (\PSQ) or gluinos (\PSg).
These particles decay directly, or via intermediate new particles,
to quarks and an LSP, where the LSP is denoted as \lsp in the following.
The signal events are generated at LO using \MADGRAPH{}5, with up to two additional partons.
The cross sections are determined at NLO and include
the resummation of soft gluon emission at the accuracy of next-to-leading-log (NLL)
calculations~\cite{bib-nlo-nll-01,bib-nlo-nll-02,bib-nlo-nll-03,bib-nlo-nll-04,bib-nlo-nll-05,SMSXsecUnc}.
Both for the generation of signal events and the calculation of
\PSQ (\PSg) production cross section, the contribution of \PSg (\PSQ) production
is effectively removed by assuming the gluino (squark) mass to be very large.

\begin{figure}[htbp]
 \centering
   \includegraphics[width=0.49\textwidth]{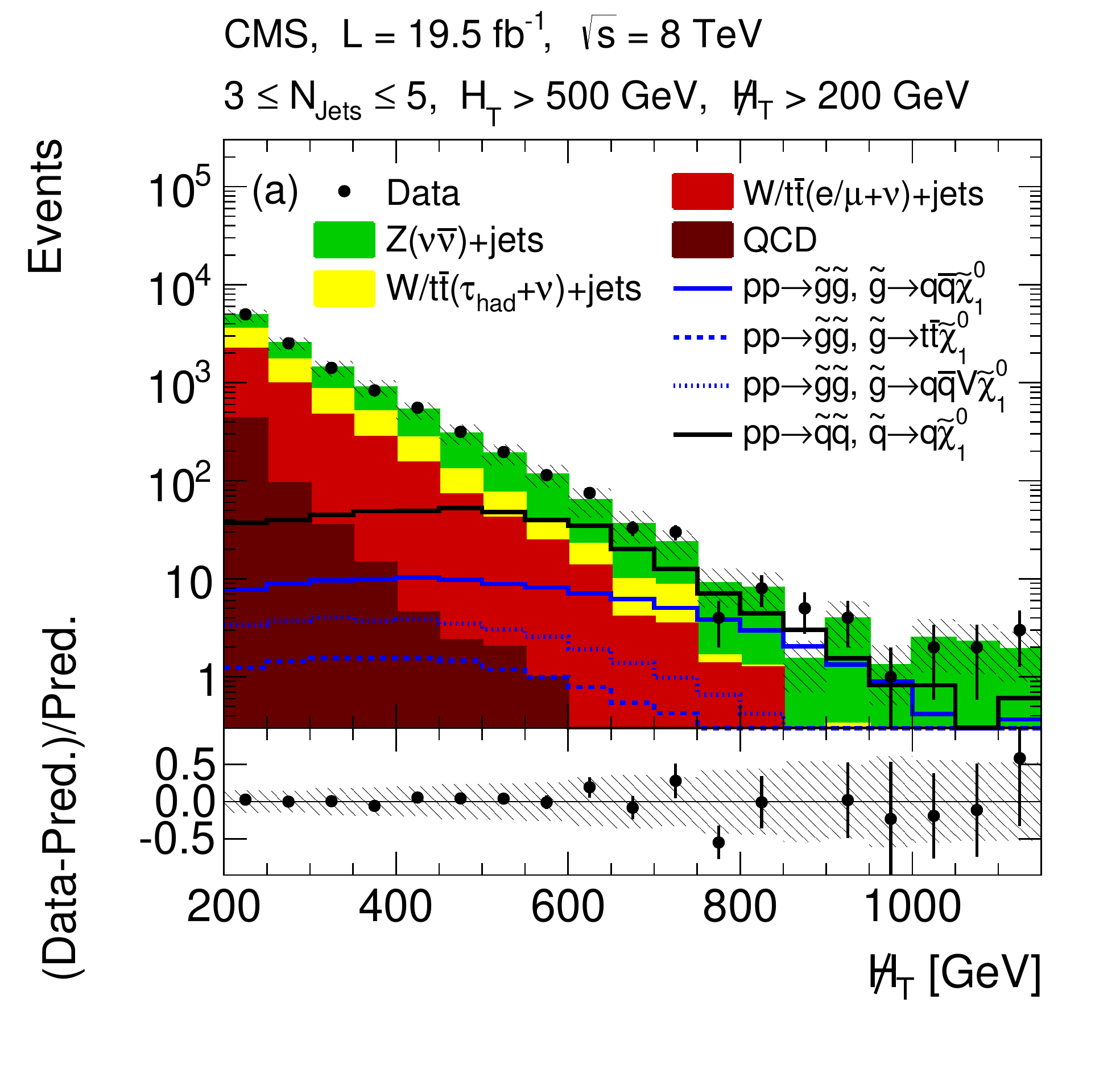}
   \includegraphics[width=0.49\textwidth]{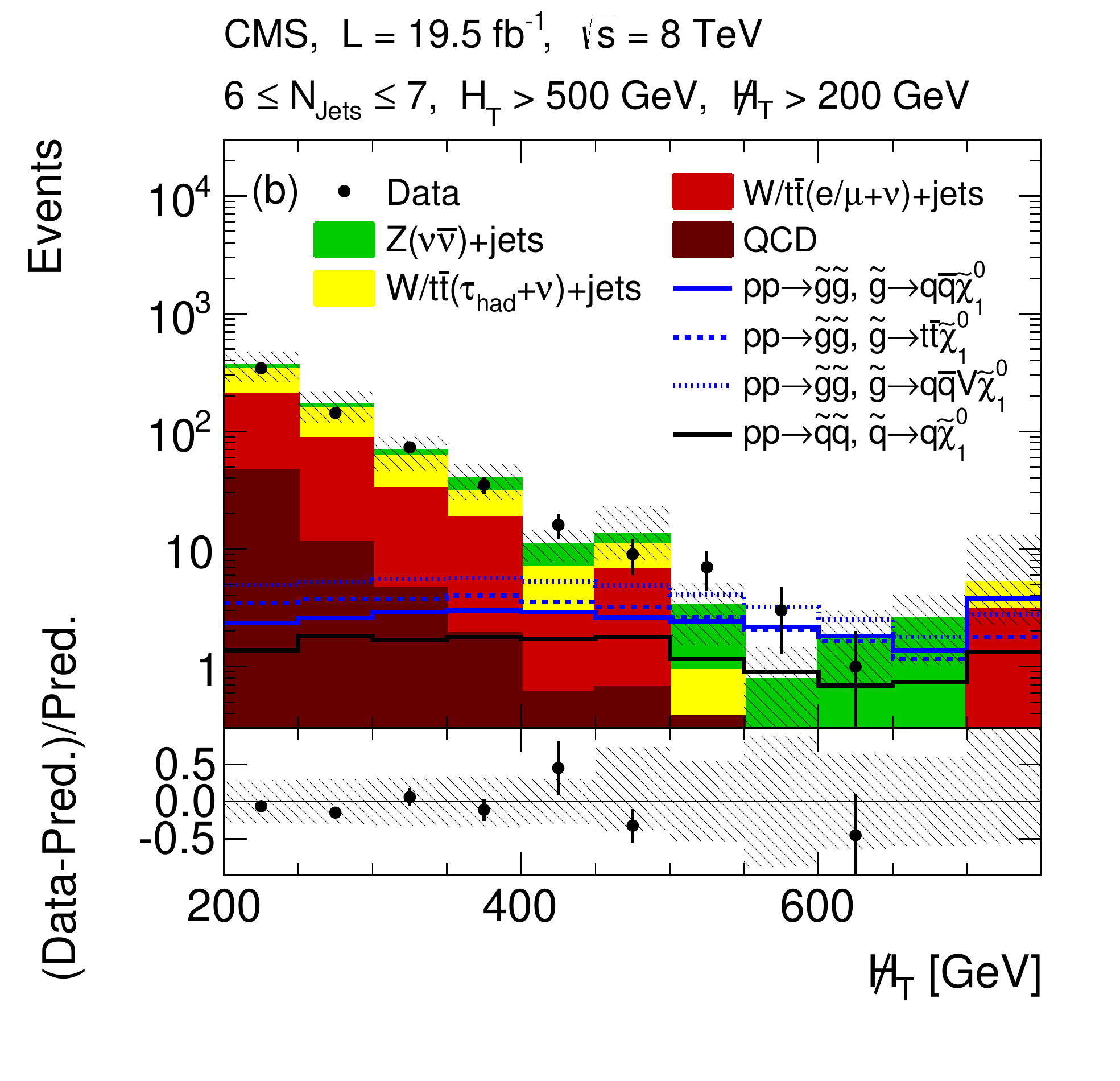}
   \includegraphics[width=0.49\textwidth]{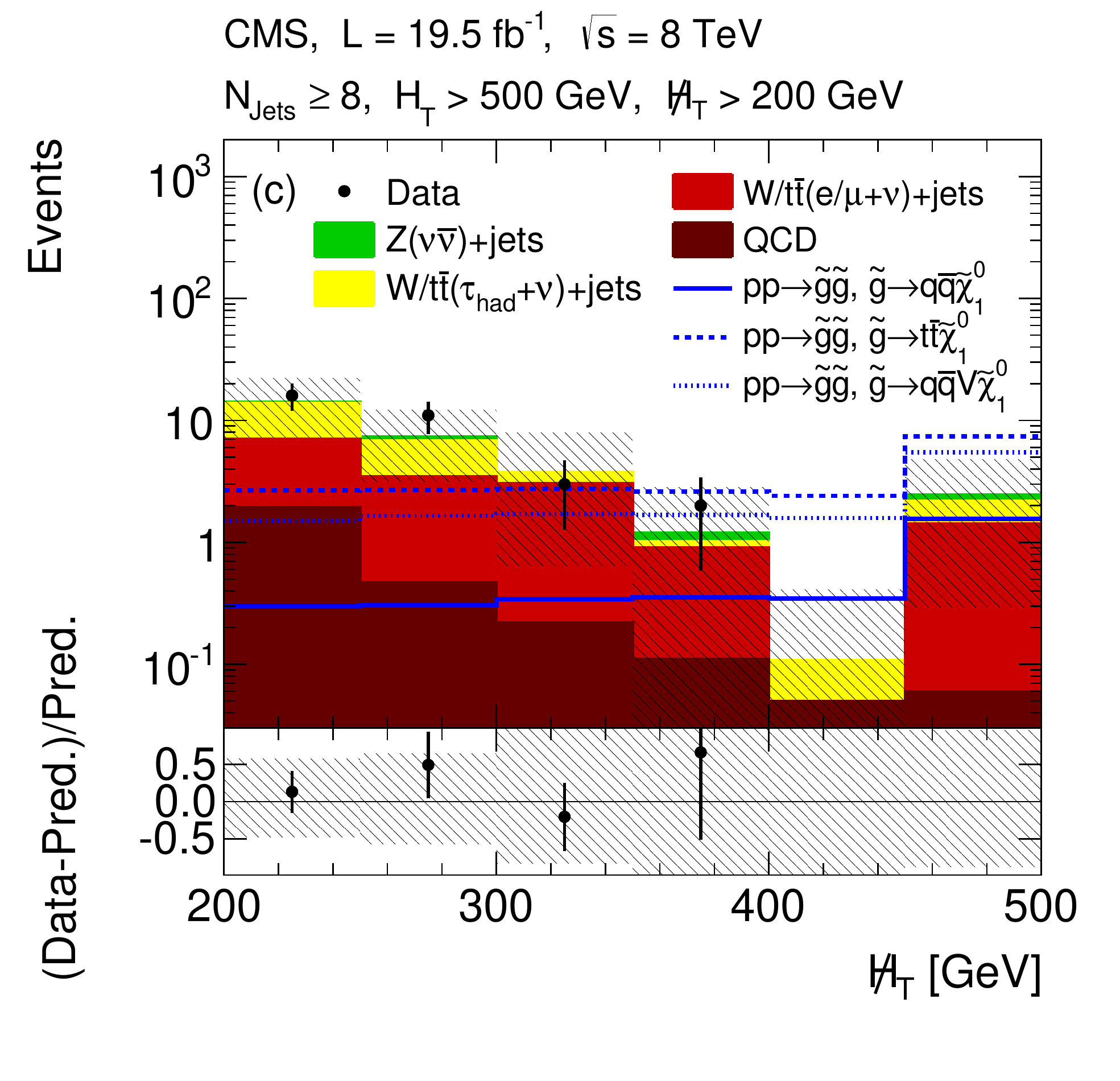}
 \caption{Observed \MHT distributions compared to the predicted backgrounds
for search regions with $\HT > 500$\GeV and jet multiplicity intervals of (a)~3--5, (b)~6--7, and (c)~$\geq$8.
The background distributions are stacked. The last bin contains the overflow. The hatched region indicates the uncertainties of the background predictions.
The ratio of data to the background is shown in the lower plots.
The \MHT distributions expected from events with
\PSg{} and \PSQ pair production, with either
$m_{\PSg} = 1.1\TeV$ and $m_{\lsp} = 125\GeV$ or
$m_{\PSQ} = 700\GeV$ and $m_{\lsp} = 100\GeV$,
are overlaid.}
 \label{fig:SummaryMHTBaselineNJets}
\end{figure}

Several decay modes of gluinos are considered here,
$\PSg\to\qqbar+\lsp$, $\PSg\to\ttbar+\lsp$,
and $\PSg\to\qqbar+\PSGcpm/\PSGc_2^0$
where $\PSGc_1^\pm\to\PW+\lsp$ and $\PSGc_2^0\to\cPZ{}+\lsp$.
The branching fraction for the different decay modes is assumed, in turn, to be 100\%,
except for the $\PSg\to\qqbar+\PSGc$ process,
where the decay proceeds via $\PSGc_1^+$, $\PSGc_1^-$
and $\PSGc_2^0$ particles with equal probability.
Squark production is studied in the decay mode $\PSQ \to \cPq + {\lsp}$.
The models are studied in the parameter space of the mass of the LSP versus the mass of
the gluino or squark.
The \MHT distributions observed for the three intervals of jet multiplicity
are shown in Fig.~\ref{fig:SummaryMHTBaselineNJets} in comparison to the SM
background prediction.
The \MHT distributions expected from gluino or squark pair production
are overlaid for $m_{\PSg}$ = 1.1 TeV and $m_{\lsp}$ = 125\GeV, and for
$m_{\PSQ}$ = 700\GeV and $m_{\lsp}$ = 100\GeV, in various decay modes.

The 95\% confidence level (CL) upper limits on the signal production cross section are set
using the LHC-style CL$_\mathrm{s}$ criterion ~\cite{aread,tjunk,ATL-PHYS-PUB-2011-011}.
The signal acceptance and
efficiencies, and corresponding uncertainties for the 36 exclusive search regions, along
with the background estimates discussed above, are combined into a likelihood that is
used to construct the test statistic based on the profile likelihood ratio.
The uncertainties of the signal acceptance and efficiency due to several sources
are taken into account when cross section upper limits are determined.
The uncertainties due to the
luminosity determination (2.6\%)~\cite{LUMI-PAS}, trigger inefficiency (2\%),
and event cleaning procedure (3\%)~\cite{METJINST} are the same for all signal
models and search regions.
The uncertainty from the measurement of the jet energy scale and jet energy resolution~\cite{JETJINST}
leads to uncertainties of 2--8\% and 1--2\% in signal acceptance.
The variation of PDFs~\cite{Botje:2011sn}
results in 1--8\% uncertainty from the signal acceptance. The rate of
initial-state radiation in the signal event simulation is corrected
to correspond to that measured in data~\cite{isrfsr}, leading to a corresponding
uncertainty of 22\% for model points with small differences between the masses
of the gluino or squark and the \lsp.
For larger mass differences, this uncertainty is typically less than a few percent.

\begin{figure}[tbph]
 \centering
   \includegraphics[width=0.49\textwidth]{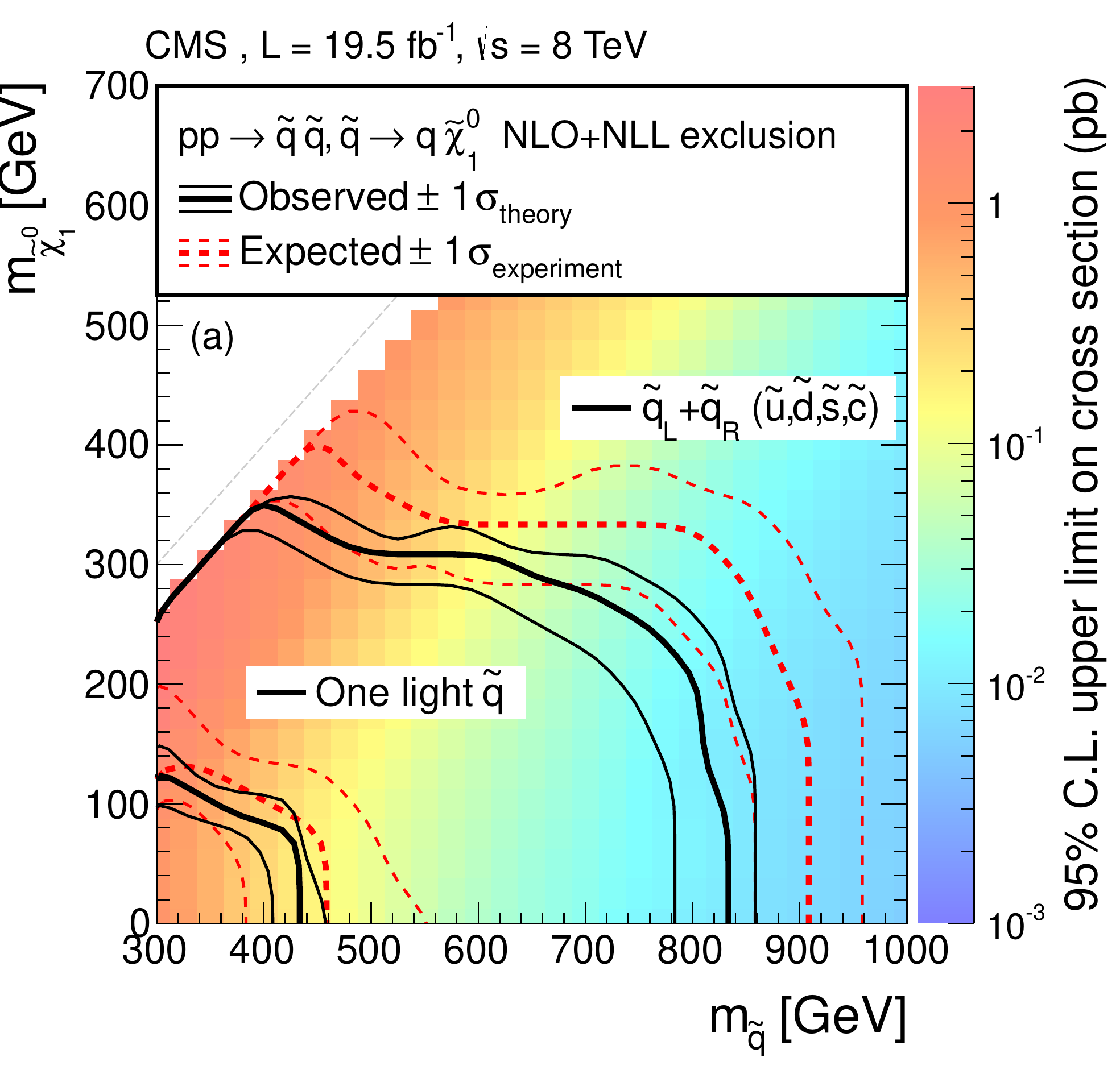}
   \includegraphics[width=0.49\textwidth]{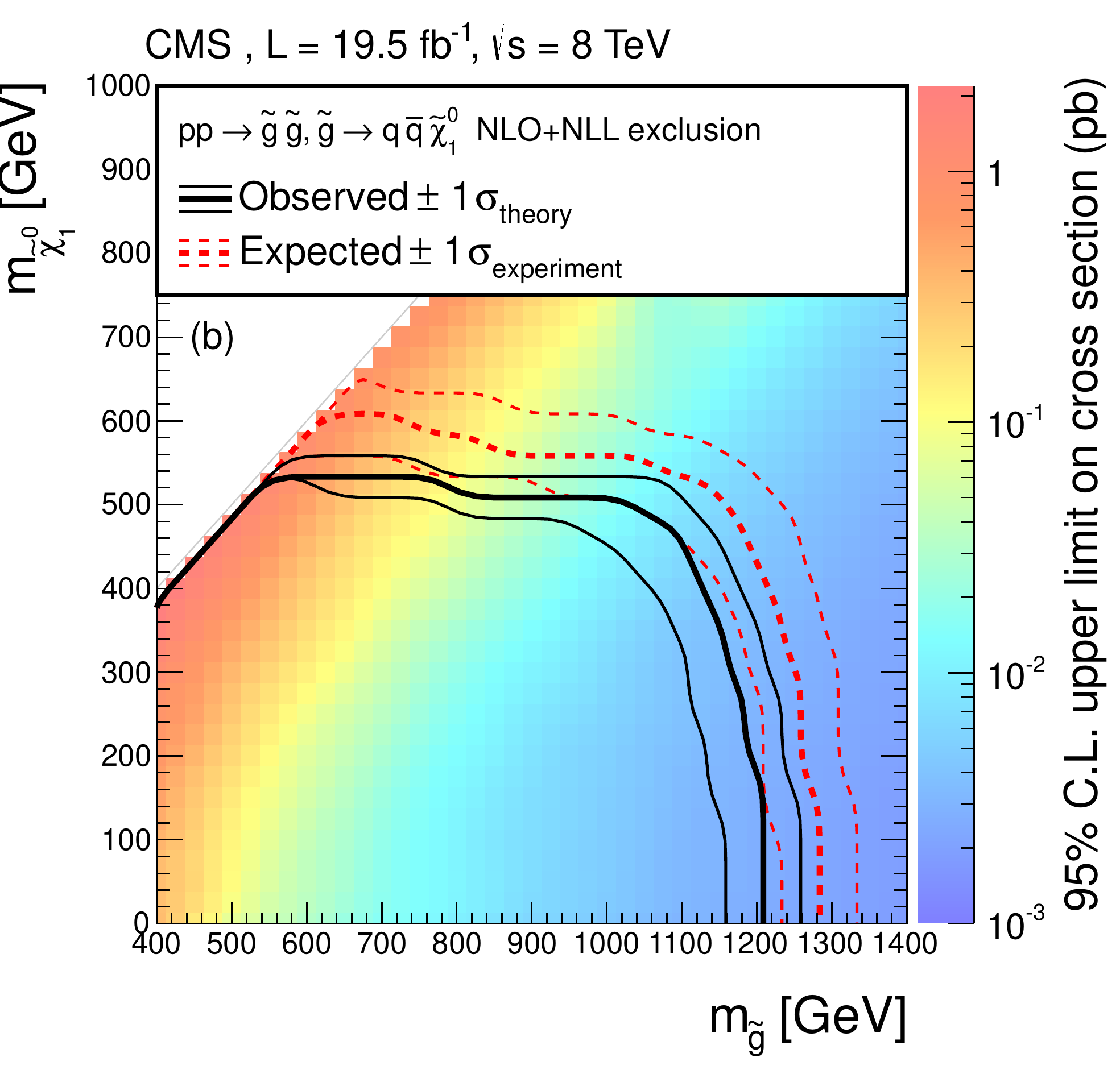}
   \includegraphics[width=0.49\textwidth]{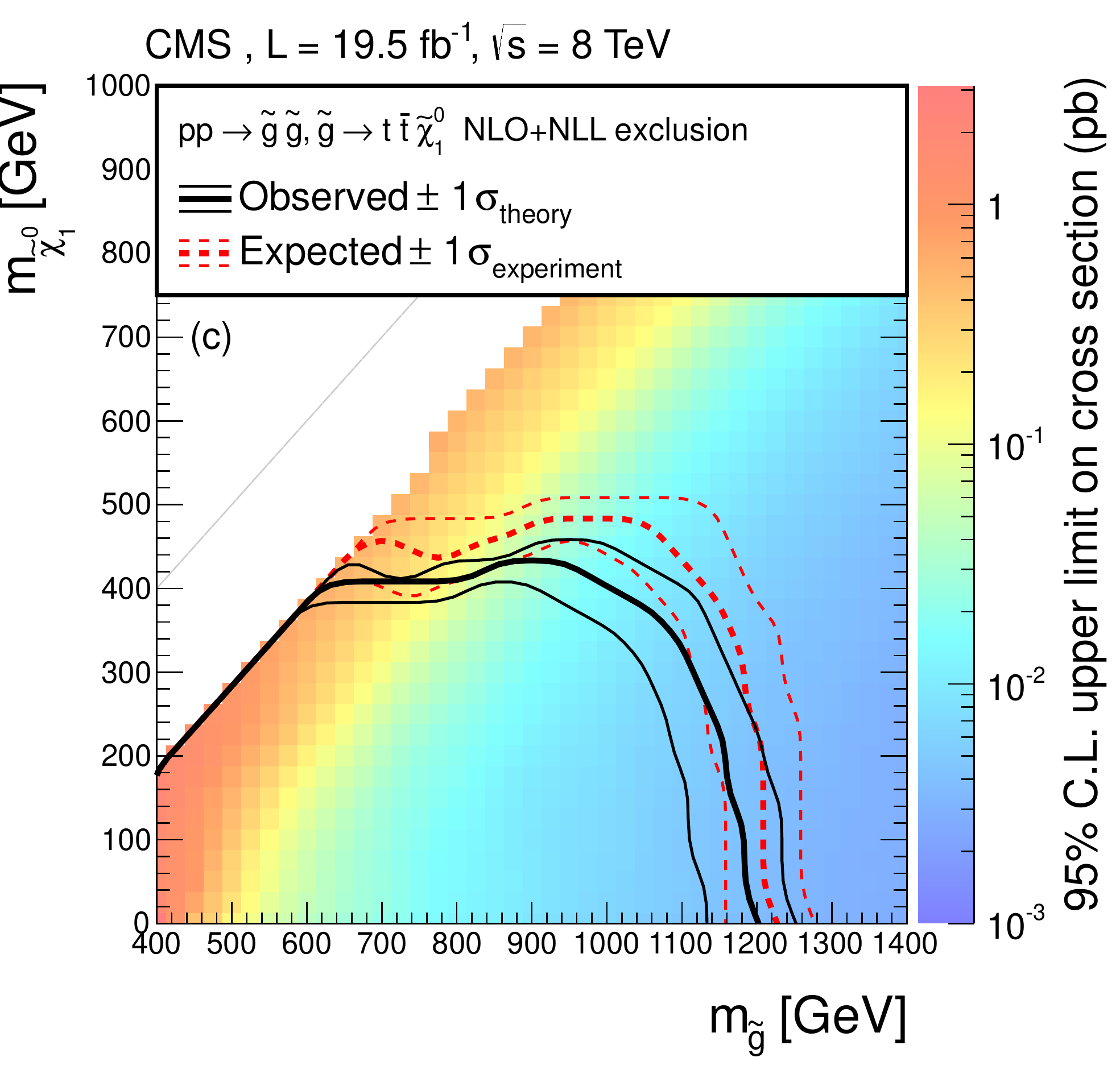}
   \includegraphics[width=0.49\textwidth]{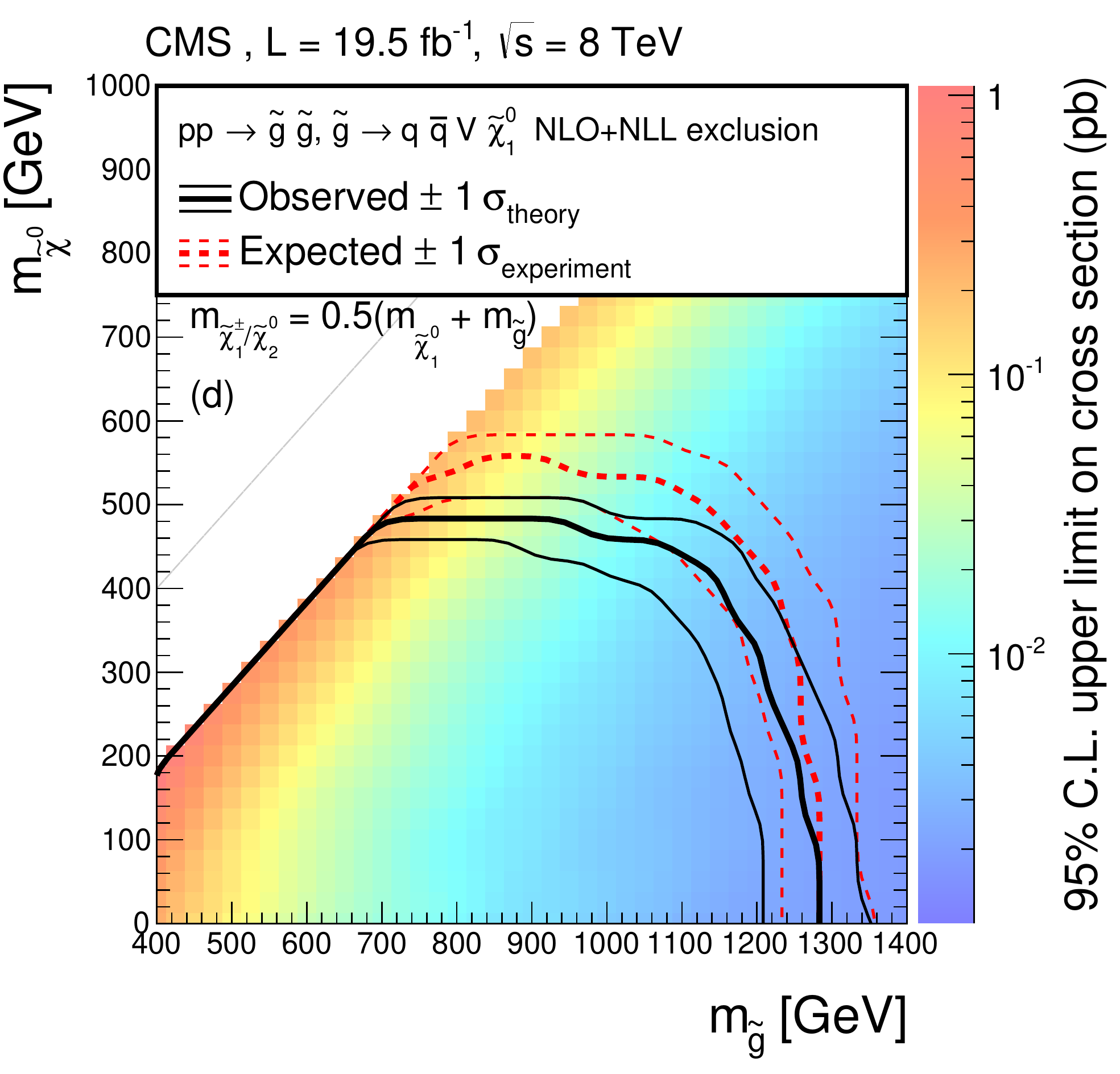}
  \caption{The observed and expected 95\% CL upper limits on the (a)~$\PSQ\PSQ$ and (b-d)~$\PSg\PSg$ production cross sections in either the ($m_{\PSQ}$, $m_{\lsp}$) or the ($m_{\PSg}$, $m_{\lsp}$) plane obtained with the simplified models. For the $\PSQ\PSQ$ production the upper set of curves corresponds to the scenario when the first two generations of squarks are degenerate and light, while the lower set corresponds to only one light accessible squark.}
  \label{fig:limitsT2qqT1qqqqT1tttt}
\end{figure}

The observed and expected $95\%$ CL upper limits on the signal cross section are shown for the production of a $\PSQ\PSQ$ pair
with $\PSQ \to \cPq + {\lsp}$ in Fig.~\ref{fig:limitsT2qqT1qqqqT1tttt}(a), a $\PSg\PSg$ pair with
 $\PSg \to \qqbar + {\lsp}$ in Fig.~\ref{fig:limitsT2qqT1qqqqT1tttt}(b),
a $\PSg\PSg$ pair with $\PSg \to \ttbar + {\lsp}$ in Fig.~\ref{fig:limitsT2qqT1qqqqT1tttt}(c),
and a $\PSg\PSg$ pair with $\PSg \to \qqbar + \PW/\cPZ + {\lsp}$ in Fig.~\ref{fig:limitsT2qqT1qqqqT1tttt}(d),
in the ($m_{\PSQ}$, $m_{\lsp}$) and ($m_{\PSg}$, $m_{\lsp}$) planes.
The contours show the exclusion regions for the signal production
cross sections obtained using the NLO+NLL calculations. The exclusion contours are also presented
when the signal cross section is varied by
changing the renormalization and factorization scales by a factor of two
and using the PDF uncertainty based on the CTEQ6.6~\cite{CTEQ66} and MSTW2008~\cite{MSTW2008} PDF sets.
Conservatively, by comparing the observed limit to the theoretical cross
section minus its one-standard-deviation uncertainty, for the cases
where the gluino decays as $\PSg \to \qqbar + {\lsp}$, $\PSg \to \ttbar + {\lsp}$,
and $\PSg \to \qqbar + \PW/\cPZ + {\lsp}$, gluino masses up to 1.16, 1.13, and 1.21 TeV are
excluded, respectively, for $m_{\lsp}<100$\GeV.
For direct $\PSQ\PSQ$ production of the first two generations of squarks
($\widetilde{\cPqu}_{L/R}$, $\widetilde{\cPqd}_{L/R}$, $\widetilde{\cPqc}_{L/R}$, $\widetilde{\cPqs}_{L/R}$),
values of $m_{\PSQ}$ below 780\GeV are excluded for $m_{\lsp}<200$\GeV. If
only one of these squarks is light, then $m_{\PSQ}$ values
below 400\GeV are excluded for $m_{\lsp}<80$\GeV. The expected search sensitivity
is improved with respect to our similar analysis~\cite{RA2_2011} based on
the 7 TeV data set by up to about 200\GeV in the values of $m_{\PSg}$, $m_{\PSQ}$
and $m_{\lsp}$.

\section{Summary}
\label{sec:conclusions}

An inclusive search for supersymmetry has been performed
in multijet events with $\njets = 3$--5, 6--7, and $\geq$8, and large
missing transverse momentum. The data sample
corresponds to an integrated luminosity of $\fulllumi\fbinv$ collected
in 8\TeV pp collisions during the year 2012
with the CMS detector at the LHC.
The analysis extends the supersymmetric parameter space explored
by searches in the all-hadronic final state.  The observed numbers
of events are found to be consistent with the expected
standard model background, which is evaluated from the data.
The results are  presented in the context of simplified
models, where final states are described by the pair production of new particles
decaying to one, two, or more jets and a weakly interacting stable neutral
particle, \eg the lightest supersymmetric particle (LSP).
Squark masses below 780\\GeV and gluino masses of up to 1.1--1.2\TeV are excluded at 95\% CL
within the studied models for LSP masses below 100\GeV.

\section*{Acknowledgments}
\label{sec:ack}
We congratulate our colleagues in the CERN accelerator departments for the excellent performance of the LHC and thank the technical and administrative staffs at CERN and at other CMS institutes for their contributions to the success of the CMS effort. In addition, we gratefully acknowledge the computing centres and personnel of the Worldwide LHC Computing Grid for delivering so effectively the computing infrastructure essential to our analyses. Finally, we acknowledge the enduring support for the construction and operation of the LHC and the CMS detector provided by the following funding agencies: BMWF and FWF (Austria); FNRS and FWO (Belgium); CNPq, CAPES, FAPERJ, and FAPESP (Brazil); MES (Bulgaria); CERN; CAS, MoST, and NSFC (China); COLCIENCIAS (Colombia); MSES and CSF (Croatia); RPF (Cyprus); MoER, SF0690030s09 and ERDF (Estonia); Academy of Finland, MEC, and HIP (Finland); CEA and CNRS/IN2P3 (France); BMBF, DFG, and HGF (Germany); GSRT (Greece); OTKA and NIH (Hungary); DAE and DST (India); IPM (Iran); SFI (Ireland); INFN (Italy); NRF and WCU (Republic of Korea); LAS (Lithuania); MOE and UM (Malaysia); CINVESTAV, CONACYT, SEP, and UASLP-FAI (Mexico); MBIE (New Zealand); PAEC (Pakistan); MSHE and NSC (Poland); FCT (Portugal); JINR (Dubna); MON, RosAtom, RAS and RFBR (Russia); MESTD (Serbia); SEIDI and CPAN (Spain); Swiss Funding Agencies (Switzerland); NSC (Taipei); ThEPCenter, IPST, STAR and NSTDA (Thailand); TUBITAK and TAEK (Turkey); NASU (Ukraine); STFC (United Kingdom); DOE and NSF (USA).

Individuals have received support from the Marie-Curie programme and the European Research Council and EPLANET (European Union); the Leventis Foundation; the A. P. Sloan Foundation; the Alexander von Humboldt Foundation; the Belgian Federal Science Policy Office; the Fonds pour la Formation \`a la Recherche dans l'Industrie et dans l'Agriculture (FRIA-Belgium); the Agentschap voor Innovatie door Wetenschap en Technologie (IWT-Belgium); the Ministry of Education, Youth and Sports (MEYS) of Czech Republic; the Council of Science and Industrial Research, India; the Compagnia di San Paolo (Torino); the HOMING PLUS programme of Foundation for Polish Science, cofinanced by EU, Regional Development Fund; and the Thalis and Aristeia programmes cofinanced by EU-ESF and the Greek NSRF.

\bibliography{auto_generated}   

\cleardoublepage \appendix\section{The CMS Collaboration \label{app:collab}}\begin{sloppypar}\hyphenpenalty=5000\widowpenalty=500\clubpenalty=5000\textbf{Yerevan Physics Institute,  Yerevan,  Armenia}\\*[0pt]
S.~Chatrchyan, V.~Khachatryan, A.M.~Sirunyan, A.~Tumasyan
\vskip\cmsinstskip
\textbf{Institut f\"{u}r Hochenergiephysik der OeAW,  Wien,  Austria}\\*[0pt]
W.~Adam, T.~Bergauer, M.~Dragicevic, J.~Er\"{o}, C.~Fabjan\cmsAuthorMark{1}, M.~Friedl, R.~Fr\"{u}hwirth\cmsAuthorMark{1}, V.M.~Ghete, C.~Hartl, N.~H\"{o}rmann, J.~Hrubec, M.~Jeitler\cmsAuthorMark{1}, W.~Kiesenhofer, V.~Kn\"{u}nz, M.~Krammer\cmsAuthorMark{1}, I.~Kr\"{a}tschmer, D.~Liko, I.~Mikulec, D.~Rabady\cmsAuthorMark{2}, B.~Rahbaran, H.~Rohringer, R.~Sch\"{o}fbeck, J.~Strauss, A.~Taurok, W.~Treberer-Treberspurg, W.~Waltenberger, C.-E.~Wulz\cmsAuthorMark{1}
% ...
\vskip\cmsinstskip
% ...
1:~~Also at Vienna University of Technology, Vienna, Austria\\
2:~~Also at CERN, European Organization for Nuclear Research, Geneva, Switzerland\\
% ...

\end{sloppypar}
\end{document}